\documentclass{pnastwo}

\usepackage[dvipdfmx]{graphicx}
\usepackage{PNAStwoF}

\usepackage{amssymb,amsfonts,amsmath,amsthm}
\newtheorem{thm}{Theorem}[]
\newtheorem{lem}[]{Lemma}
\newtheorem{cor}[]{Corollary}

\contributor{Submitted to Proceedings
of the National Academy of Sciences of the United States of America}
\url{www.pnas.org/cgi/doi/10.1073/pnas.0709640104}
\copyrightyear{2008}
\issuedate{Issue Date}
\volume{Volume}
\issuenumber{Issue Number}
\begin{document}

%%%%%%%%%%%%%%%%%%%%%%%%%%%%%%

%% For titles, only capitalize the first letter
%% \title{Almost sharp fronts for the surface quasi-geostrophic equation}

\title{Characterizing Multivariate Information Flows}

%% Enter authors via the \author command.  
%% Use \affil to define affiliations.
%% (Leave no spaces between author name and \affil command)

%% Note that the \thanks{} command has been disabled in favor of
%% a generic, reserved space for PNAS publication footnotes.

%% \author{<author name>
%% \affil{<number>}{<Institution>}} One number for each institution.
%% The same number should be used for authors that
%% are affiliated with the same institution, after the first time
%% only the number is needed, ie, \affil{number}{text}, \affil{number}{}
%% Then, before last author ...
%% \and
%% \author{<author name>
%% \affil{<number>}{}}

% \author{<author name>
%\affil{<number>}{<Institution>}} One number for each institution.
% The same number should be used for authors that
% are affiliated with the same institution, after the first time
% only the number is needed, ie, \affil{number}{text}, \affil{number}{}
% Then, before last author ...
% \and
% \author{<author name>
% \affil{<number>}{}}

%% For example, assuming Garcia and Sonnery are both affiliated jwith
%% Universidad de Murcia:
%% \author{Roberta Graff\affil{1}{University of Cambridge, Cambridge,
%% United Kingdom},
%% Javier de Ruiz Garcia\affil{2}{Universidad de Murcia, Bioquimica y Biologia
%% Molecular, Murcia, Spain}, \and Franklin Sonnery\affil{2}{}}

 \author{Shohei Hidaka\affil{1}{Japan Advanced Institute of Science and Technology, 1-1 Asahidai, Nomi, Ishikawa,
 Japan}}

%\author{a}

\contributor{Submitted to Proceedings of the National Academy of Sciences
of the United States of America}

%% The \maketitle command is necessary to build the title page.
\maketitle

%%%%%%%%%%%%%%%%%%%%%%%%%%%%%%%%%%%%%%%%%%%%%%%%%%%%%%%%%%%%%%%%
\begin{article}

\begin{abstract} 
One of the crucial steps in scientific studies is to specify dependent relationships among factors in a system of interest. Given little knowledge of a system, can we characterize the underlying dependent relationships through observation of its temporal behaviors? In multivariate systems, there are potentially many possible dependent structures confusable with each other, and it may cause false detection of illusory dependency between unrelated factors. The present study proposes a new information-theoretic measure with consideration to such potential multivariate relationships. The proposed measure, called multivariate transfer entropy, is an extension of transfer entropy, a measure of temporal predictability. In the simulations and empirical studies, we demonstrated that the proposed measure characterized the latent dependent relationships in unknown dynamical systems more accurately than its alternative measure.
\end{abstract}

%% When adding keywords, separate each term with a straight line: |
%\keywords{term | term | term}

%% Optional for entering abbreviations, separate the abbreviation from
%% its definition with a comma, separate each pair with a semicolon:
%% for example:
%% \abbreviations{SAM, self-assembled monolayer; OTS,
%% octadecyltrichlorosilane}

% \abbreviations{}

%% The first letter of the article should be drop cap: \dropcap{}
%\dropcap{I}n this article we study the evolution of ''almost-sharp'' fronts

%% Enter the text of your article beginning here and ending before
%% \begin{acknowledgements}
%% Section head commands for your reference:
%% \section{}
%% \subsection{}
%% \subsubsection{}

%-- text of paper here --

%\section{Characterization of a system from an observed time series}
\dropcap{O}ne of crucial steps in scientific studies is the characterization of a system of interest - specification of dependent relationships among factors or subcomponents in the system \cite{Gershenfeld1993}. The characterization is often the early stage of analysis before proceeding to a more specific description of the systems, and it requires little or no prior knowledge of the underlying mechanism of the system. In the present study, we propose a measure which may serve as such an early-stage characterization of dependency in a multivariate system with little knowledge through observation of its temporal behaviors. 

A basic problem concerned in the present study is how we can quantify and detect dependency between each pair of variables in a multivariate system. In a multivariate system, a variable X generated by a stochastic or a deterministic process is said to be conditionally dependent of variable $Y$ given $Z$ if its future state of $X$ is partially determined by another variable $Y$ given $Z$. In particular, we consider temporal behaviors of a system in which we measure dependency from a set of $N$ variables at time $t$ to the set of $N$ variables at time $t+1$. In this formulation, the temporal dependency of a system is characterized with $N^2$ conditional dependency between $X_t$ and $Y_{t+1}$ given $Z_{t}$.

In a linear system, which can be decomposed into separable subcomponents without interaction among them, characterization of temporal dependency is straightforward. For stationary linear processes, auto- and cross-correlation sufficiently characterizes a set of its linear properties. In contrast, a nonlinear system with or without a stochastic component, which is not decomposable to subcomponents, needs to be characterized with consideration to its interaction among subcomponents. A set of information theoretic measures has been proposed as a nonlinear counterpart of auto- and cross-correlation for a nonlinear system \cite{Kantz1997}.
One of such information-theoretic measures, transfer entropy, has found a wide range of applications and has successfully characterized many empirical systems \cite{Schreiber2000}.
In the present study, we propose an extended information-theoretic measure, called {\it multivariate transfer entropy} (MTE), for characterization of multivariate dependency. The proposed measure is a natural generalization of transfer entropy, and it concerns potential confounding relationships among three or more variables in multivariate temporal dynamics. Thus we illustrate the extended measure after a brief overview of the related development of the information theory.

\subsection{Information Theory}

Ever since its establishment by Shannon \cite{Shannon1948}, information theory has played a crucial role in mathematical modeling of communication.
His original formulation concerns a unidirectional information transmission through a noisy channel. In the formulation, a message $X$ generated by an information source is sent to a receiver. The receiver receives it as a message $Y$ through a stochastic channel in which the original message $X$ may be changed to $Y$ by a certain chance. This unidirectional information transmission is described with the mathematical concept {\it entropy} and {\it mutual information} of probabilistic distribution of the sent and received messages $X$ and $Y$. Entropy quantifies the amount of stochastic uncertainty of an information source by the length of codes encoding the message generated by the information source. Mutual information quantifies the relative difference between uncertainty in the two ways of coding, the random variable $Y$ alone and the variable $Y$ with additional knowledge of another variable $X$. The mutual information is maximized when $X=Y$ (noiseless channel), and it is zero at the minimum when random variables $X$ and $Y$ are independent. Thus, mutual information characterizes the properties of the noisy channel between $X$ and $Y$. 
%Mutual information gives a mathematical ground for communication theory which concerns to design an optimal information channel given constraints. 
Mutual information gives a mathematical ground for communication theory concerning the design of an optimal information channel given constraints.
Introducing the concepts of entropy, mutual information and its variants, information theory covers and connects a wide range of fields and problems such as nonlinear dynamical system, thermodyamics, electrical engineering, probability theory, statistics, mathematics, economics, computer science, and philosophy of science \cite{CoverThomas1991}.

 Despite the mathematical elegance and many successful applications, in 1973 Shannon had pointed out the theoretical limitation of the unidirectional information transmission, and gave a prospect for an extension to information theory {\it with feedback} \cite{Massey1990}. Indeed in the very same year, Marko \cite{Marko1973} proposed the extended bidirectional information network as suggested by Shannon. In his bidirectional information network, two information sources send and receive information, and its effficiency of communication is characterized with loss of information from the bidirectional information transfer. Although both Shannon and Marko highlighted the importance of bidirectional information, it has not been well-recognized in the fields until 1990s \cite{Massey1990}. More recently, Kantz and Schreiber \cite{Kantz1997,Schreiber2000} have reintroduced a {\it directed} measure of statistical dependency, called transfer entropy, which is a subset of Marko's bidirectional information theory. One of the major advantages of the transfer entropy or bidirectional information transmission over Shannon's unidirectional measure is that it enables us to distinguish which factor leads or follows another separately from the other direction. After the reintroduction of transfer entropy, it has found applications in various research fields. Such successful applications include not only engineering fields relevant to information theory but also various kinds of scientiffic research: detecting directed dependency in 
cellular automata \cite{Lizier2008}, 
machine learning \cite{Quinn2010},
chemical process \cite{Bauer2007}, 
health monitoring \cite{Nichols2005}, 
analysis of brain activity \cite{Rubinov2010,Vicente2011},  
stock markets \cite{Baek2005,Marschinski2002}, 
ecological monitoring programs \cite{Moniz2007},
music analysis \cite{Kulp2009}, 
and human-human/robot communication \cite{Hidaka2010,Hidaka2010b,Sumioka2007}. 

\subsection{Limitation of the transfer entropy}
Despite its successful applications, a potential limitation of transfer entropy has not been well recognized. A naive application of transfer entropy to three or more variables may cause an inaccurate characterization of a system. In the present study, we demonstrated this limitation, and we propose multivariate transfer entropy (MTE), a further extension of Marko's bidirectional information theory, as a solution.
%as a further extension of Marko's bidirectional information theory to three or more information sources and destinations. 
The MTE is concerned with the potential confounding relationship among three or more variables which transfer entropy does not count. Thus it extends its usability to more general situations with arbitrary topological structure of dependency among $N$ variables. In this regard, the MTE is a nonlinear analogue of partial correlation which cancels linear confounding effects of other than correlation between the focal paired variables. In order to explicitly distinguish from the MTE, hereafter we refer to the original one as { \it pairwise} transfer entropy (PTE), a special case for a bivariate system.
In the following section, we give a formal description of mutual information, transfer entropy and their relationship.

\section{Mutual information and transfer entropy}
%Shannon's original formulation concerns a unidirectional information transmission through a noisy channel. 
Consider a unidirectional information transmission through a noisy channel in which a message $X$ generated by an information source is sent to a receiver. 
%The receiver receives it as a message $Y$ through a stochastic channel in which the original message $X$ may be changed to $Y$ by a certain chance. This unidirectional information transmission is described with entropy and mutual information defined as follows. 
Let $p(x_i)$ be the probabilistic distribution of a message $X = x_i  \in \mathcal{M} = \{1,2, ..M\}$, where $\mathcal{M}$ is the set of $M$ alphabets, and a series of messages is drawn from the probabilistic distribution by the information source. Then entropy, $-\sum_{i}p(x_i)\log p(x_i)$, gives the asymptotic minimum average code length by assigning the code with length $-\log p(x_i)$ to a infinitely long series of messages $\{X_1, X_2, \hdots \}$.

Then suppose that we assign a code set $Q$ of length $-\log q(x_i)$, instead of the minimum code set $P$ of length $-\log p(x_i)$, to the message $x_i$ with probability $p(x_i)$. Its average relative difference between the code length of $Q$ and $P$ called relative entropy or Kullback-Liebler divergence, $D(P||Q) = -\sum_{i}p(x_i)( \log q(x_i) - \log p(x_i) )$, can be treated as the amount of coding error or the difference in stochastic uncertainty of $q(X)$ relative to $p(X)$. In the unidirectional information transmission described above, the entropy $H(Y)$ quantifies the amount of stochastic uncertainty of the received message $Y$. Likewise, the conditional entropy, $H(Y|X) = -\sum_{i,j}p(x_i, y_j)\log p(y_j | x_i)$, quantifies that of $Y$ on knowledge of $X$. Then mutual information $I(X; Y) = H(H)-H(Y|X)$ is defined as difference of the entropy of $Y$ relative to conditional entropy $H(Y|X)$ or symmetrically that of the entropy $X$ relative to conditional entropy $H(X|Y)$. Mutual information can be interpreted as the amount of information gain by obtaining shorter code length $H(Y|X)$, $Y$ with additional knowledge of $X$ relative to the code length $H(Y)$ of $Y$ alone.

\subsection{Transfer entropy and bidirectional information network}
\begin{figure}[t,clip]
   \begin{center}
\includegraphics[width=1\linewidth, clip]{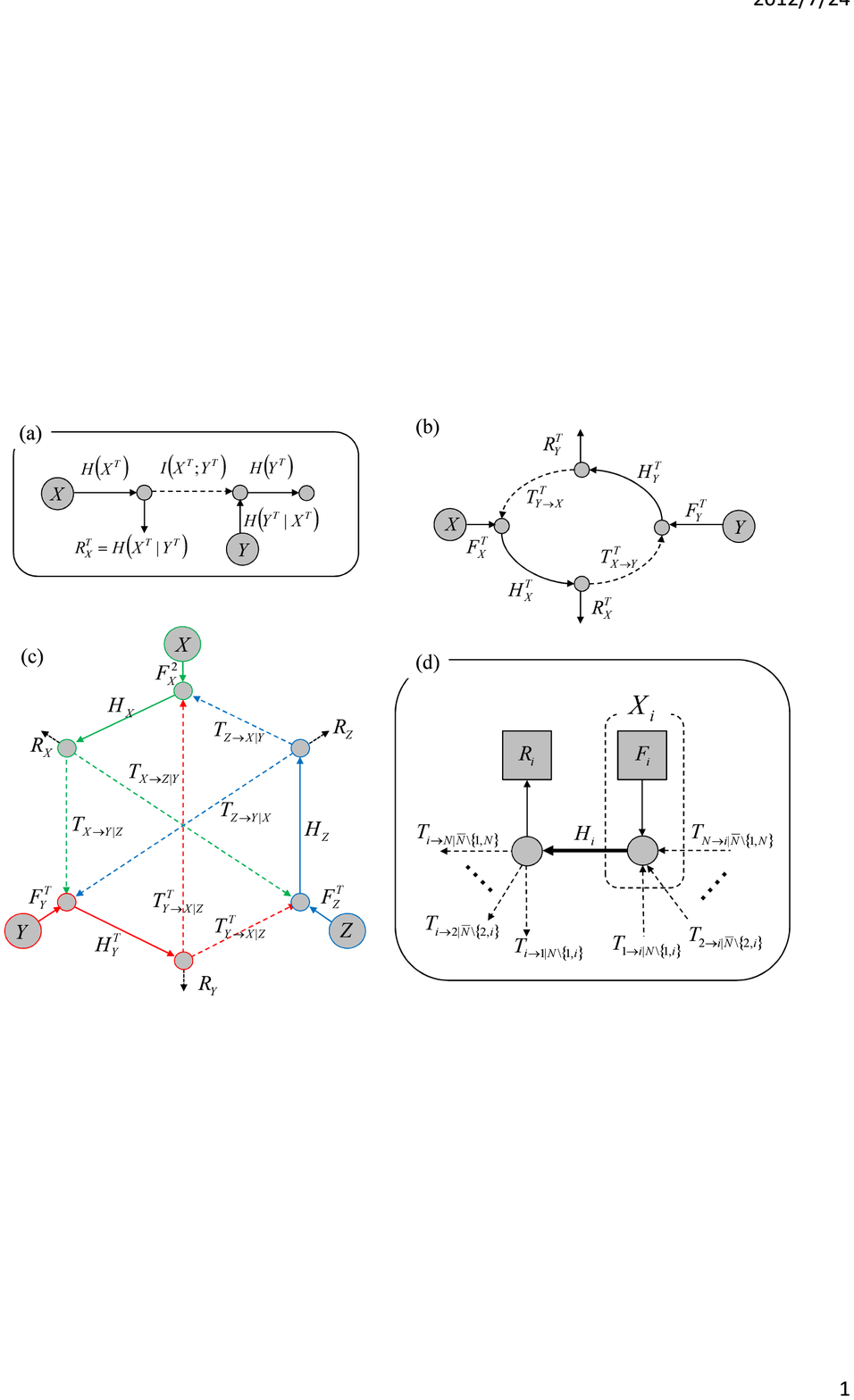}
    \caption{(a) Unidirectional and (b) bidirectional information transmission between two variables X and Y. (c) The extended bidirectional information transmission among three variables, and (d) local in-coming and out-going information flows in a part of a general N-variable bidirectional information network.\label{fig-MStransfer_entropy}}
    \end{center}
  \end{figure}

%[Neeraj] [You should make it clear that there is no noise, as the notation is similar to the unidirectional channel with noise situation you described above.]
Marko \cite{Marko1973} has given a reinterpretation of the Shannon's unidirectional transmission as a network of information flows, and showed its bidirectional extension. Figure \ref{fig-MStransfer_entropy}a depicts the unidirectional information transmission interpreted as a network flows \footnote{In the network, the superscript $T$ is a set of time indices, and $X^T = {X^1, X^2, \hdots, X^T}$ is a group of variables in which we obtain the Shannon's entropy by identifying $X = X^T$.}. In the network, the mutual information $I(X^T; Y^T)$, as some amount of entropy of the information source $X^T$, is flown into the $Y^T$. The entropy of received message $Y^T$ is the sum of the in-coming flow $I(X^T; Y^T)$ and the uncertainty of $Y^T$ alone without $X^T$, the conditional entropy $H( Y^T | X^T )$.

This unidirectional information network is a special case of the bidirectional network (Figure \ref{fig-MStransfer_entropy}b). We follow Marko's terminological conventions except for the term entropy rate and (pairwise) transfer entropy, which have become more standard after \cite{CoverThomas1991,Schreiber2000}. Unlike the one originally proposed by \cite{Marko1973}, the following formulation needs to assume neither stationary nor Markovity of time series in theory \footnote{However, estimation of transfer entropy often requires these properties due to finite sample size of dataset in practice.
Marko \cite{Marko1973} considered the limit of infinite long time series, which we can obtain by making $T \rightarrow \infty$ in the present formulation of time series of a finite length. Specifically,  in this paper, we work simply with the quantities $H_X^T$, 
since its limit is not essential in to the arguments presented here.}. 

Suppose we have two series of random variables $\bar{X}^{T} = \{ X^{1}, X^{2}, \hdots, X^{T}\}$ and $\bar{Y}^{T} = \{ Y^{1}, Y^{2}, \hdots, Y^{T} \}$  over discrete time $t = 1, 2, \hdots , T$ where the top bar $\bar{X}^{T}$ means the set of random variables with superscript specifying time indices from time $1$ to time $T$. As in the unidirectional communication, we start with a measure of uncertainty in a single variable. {\it Entropy rate} $H_X^T$ is the sum of conditional entropies of $X^t$ given its past states  $\bar{X}^{t-1}$ $(t = 1, 2, \hdots, T-1 )$ \cite{CoverThomas1991}. 
\begin{equation}
 H_{X}^{T} \equiv \sum_{t = 1}^{T} H(X^{t}| \bar{X}^{t-1})
\end{equation}
where $ H(X^{t} | \emptyset ) = H( X^{t} )$ and $X^t = \emptyset$ for $t < 1$.
The entropy rate $H_X^T/ T$ is the average increase at each step in the entropy of variable $X$ by normalizing with length of time series $T$.
%Similarly, we define $H_{Y}^{T} \equiv \sum_{t = 1}^{T} H(Y^{t}| \bar{Y}^{t-1})$.
Similarly, the sum of uncertainties of a random variable $X$ at time $t$ conditioned on knowledge of the past states of $\{\bar{X}^{t-1}, \bar{Y}^{t-1}\}$ for $t = 1, 2, \hdots, T$ is called {\it free entropy} $F_{X}^{T}$.  Formally, we define as follows.
\begin{equation}
 F_{X}^{T} = \sum_{t=1}^{T}H(X^{t} | \bar{X}^{t-1}, \bar{Y}^{t-1} )
\end{equation}
Similarly, we write $ F_{Y}^{T} = \sum_{t=1}^{T}H(Y^{t} | \bar{X}^{t-1}, \bar{Y}^{t-1} )$. 
%the bidirectional model consider $H( x^{t} | \bar{X}^{t-1}, )$
The pairwise transfer entropy from $Y$ to $X$ at time $T$ is defined as the sum of reductions in uncertainty of $X^{t}$ conditional on knowledge of the past states of two variables $\{ \bar{X}^{t-1}, \bar{Y}^{t-1} \}$ for $t=1, 2, \hdots, T$.
%which is called (pairwise) { \it transfer entropy}.
\begin{equation}
 %T_{Y \rightarrow X} = I(x^{t}; \bar{Y}^{t-1} | \bar{X}^{t-1}) = H_{X} - \sum_{t=1}^{T}H(x^{t} | \bar{X}^{t-1}, \bar{Y}^{t-1} )
 T_{Y \rightarrow X}^{T} = H_{X}^{T} - F_{X}^{T} %\sum_{t=1}^{T}H(x^{t} | \bar{X}^{t-1}, \bar{Y}^{t-1} ) 
                     = \sum_{t=1}^{T}I(X^{t}; \bar{Y}^{t-1} | \bar{X}^{t-1})
\label{eq-PairwiseTransferEntropy}
\end{equation} 
where $I(X; Y | Z) = H(X | Z) - H( X| Y, Z)$ is conditional mutual information between $X$ and $Y$ given $Z$,
and $I(X; Y | \emptyset ) = I( X; Y )$.
Similarly, $T_{X \rightarrow Y}^{T} = H_{Y} - F_{Y}$, and $T_{X \rightarrow Y}^{T} \neq T_{Y \rightarrow X}^{T}$ in general. 
Transfer entropy can be interpreted as directed ``information transmission'' from $Y$ to $X$, since 
$H( X^{t} | \bar{X}^{t-1} ) \ge H( X^{t} | \bar{X}^{t-1}, \bar{Y}^{t-1} )$ and $H( Y^{t} | \bar{X}^{t-1} ) \ge H( Y^{t} | \bar{X}^{t-1}, \bar{Y}^{t-1} )$ if and only if the series of variable $\{\bar{X}^{t}\}$ is independent of the past states of another variable $\{\bar{Y}^{t-1}\}$ for $t=1, 2, \hdots, T$.
%dependency between variable $X$ at $t$ and the past states of $Y^{t-1}$ in which reduction of uncertainty by knowing the past states of $X^{t-1}$ is discounted.

%The second problem is information transmission in which we send a message via a channel with certain level of noise. Suppose $X$ is probabilistic variable of the message to be sent, and $Y$ is that of message sent through the channel. Conditional entropy $H(X|Y) = -\sum_{i,j}p_{i,j}(\log_{2} p_{i,j} - \log_{2} p_{j})$, in which $p_{i,j}$ is joint probability of variable $\{X, Y\}$, is uncertainty of $X$ conditional on knowledge of $Y$. Reduction in uncertainty $I(X;Y) = H(X)-H(X|Y)$ is { \it mutual information } which quantifies how much information is shared in the two variables $X$ and $Y$. Mutual information characterize property of noise in the information channel and corresponds with the maximum rate of information which we can send through the channel.

\subsection{Network properties of transfer entropy}
Marko \cite{Marko1973} has pointed out that the relationship between entropy rate, free entropy, and transfer entropy can be viewed as a bidirectional information network (Figure \ref{fig-MStransfer_entropy}b). The bidirectional network has two variables $X$ and $Y$ which send and receive messages between them. Each directed edge in the network reflects an information flow with non-negative value of corresponding entropy rate (solid line), free entropy (solid line) or transfer entropy (broken line). In each node, the total amount of in-coming information flows is identical to the total amount of out-going ones (Kirchhoff's current law). 
%Each node of the network, are connected with a directed edge. 
The entropy rate  $H_X^{T}$ is the sum of free entropy $F_X^T$ (new information at $T$) and transfer entropy $T_{Y\rightarrow X}^{T}$  (information from another variable the past states up to $T - 1$). A certain part $T_{X \rightarrow Y}^{T}$ of entropy rate $H_{X}^{T}$ is transferred to $Y$, and the rest, called {\it residual entropy} $R_{X}^{T} = H_{X}^{T} - T_{X \rightarrow Y}^{T}$, is flown out of the network. Similarly, information is transferred from the variable $Y$ to $X$.
 At each node and edge in the bidirectional network for the two variable $X$ and $Y$, two properties, non-negativity of information flows and Kirchhoff's current law, are held. We refer these to two properties as network constraints. In order for all the information flows to be non-negative, it needs to satisfy the following inequality: $R_{X}^{T} = H_{X}^{T}- T_{X\rightarrow Y}^{T} \ge 0$  and symmetrically $R_{Y}^{T} = H_{Y}^{T}- T_{Y\rightarrow X}^{T} \ge 0$. In [7], an even better inequality as follows has been suggested without a proof.
\begin{equation}
 \min( H_{X}^{t}, H_{Y}^{t}) \ge T_{X\rightarrow Y}^t+T_{Y\rightarrow X}^t
  \label{eq-MarkoInequality}
\end{equation}
We will prove a more general version of this inequality for $N$-variable system ($N>2$) in this study.

%\subsection{Transfer entropy as decomposition of total correlatio}n
\subsection{Transfer entropy as decomposition of mutual information}
Another property of the PTE is as a partial factor decomposing mutual information with certain residuals \footnote{
This relationship between {\it transfer entropy} and mutual information has been pointed out originally in \cite{Marko1973} without 
the residual term $R_{X,Y}$.
In \cite{Massey1990}, the transfer entropy $T_{X \rightarrow Y}$ was defined as $T_{X \rightarrow Y} + R_{X,Y}$ in the current notations.
%In this argument, he claimed that two information sources $X$ and $Y$ in the network is independent if the transfer entropy from $X$ to $Y$ and that from $Y$ to $X$ are both zero (Figure \ref{fig-MStransfer_entropy}). 
%However his argument is incomplete, because the following identity holds (see also inequality \ref{eq-MarkoInequality}).
}.
\begin{eqnarray}
 I( \bar{X}^{T}; \bar{Y}^{T} ) &=& T_{X \rightarrow Y} + T_{ Y \rightarrow X } + R_{X,Y}
  \label{eq-IdentityBivariateTransferEntropy}
\end{eqnarray}
where $
R_{ X, Y } = \sum_{i=1}^{N}I( x_{i}; y_{i} | \bar{X}_{i-1}, \bar{Y}_{i-1} ) \ge 0$ is non-negative due to non-negativity of conditional mutual information.
Equation \ref{eq-IdentityBivariateTransferEntropy} explicitly states PTE is an extension of mutual information
which is a special case without feedback $T_{Y\rightarrow X} + R_{X,Y} = 0$ or $T_{X\rightarrow Y} + R_{X,Y} = 0$.
%\end{eqnarray}
%We call $R_{X, Y}$ {\it residual entropy} which is difference between mutual information and sum of transfer entropies in two directions, and this has not been discussed explicitly in \cite{Marko1973}.
%Therefore, precisely speaking, two information sources are independent --or zero mutual information between them, if and only if transfer entropy in the two directions {\it and} residual entropy are all zero.

\section{ Multivariate bidirectional information network }
Here we outline an extension of the pairwise transfer entropy to more general cases with three or more information sources. Technically, there are many potentially possible multivariate extensions of PTE. However, the proposed extension is justified not just by applicability to multivariate dependency but also by holding the two properties as well as PTE.
%: the network properties and the decomposition of an extension of mutual information. 
 Analogous to PTE decomposing of mutual information, MTE decomposes total correlation, a multivariate extension of mutual information \cite{Watanabe1960} (also called multivariate constraint \cite{Garner1962} or multiinformation \cite{Studeny1999}).
Also we can view MTE as a part of bidirectional information network among $N$ variables with non-negative flow holding Kirchhoff's current law. In the following sections, we formulate a multivariate information network and overview its theoretical properties. See also Supplemental Information for the more detailed description and the mathematical proofs of the theorems.

\subsection{Formulation of multivariate information network}
In a generalized information network with $N$ variables, each variable is associated with the two nodes -- in-coming and out-going node (Figure \ref{fig-MStransfer_entropy}d).
The in-coming and out-going node of variable $i$ respectively receives and sends information from all the variables but the variable $i$.
Information flow in between the in-coming and out-going node of variable $i$ is the entropy rate of variable and free entropy of variable $i$. At the out-going node, there is some amount of information lost without being transferred to the other variables which is called residual entropy.
A special case of the information network for a three variable system is shown in Figure \ref{fig-MStransfer_entropy}c.

Let $\bar{X}_{ \mathcal{N} }^{ \mathcal{T} }$ be a set of $N \times T$ random variables indexed with the index set $\mathcal{N} = \{ 1, 2, \hdots, N \}$ and the index set for time $\mathcal{T} = \{1, 2, \hdots, T\}$.
Then let us denote
$\bar{X}_{ \mathcal{N} }^{ \mathcal{T} } = \{ \bar{X}_{1}^{ \mathcal{T} }, \bar{X}_{2}^{ \mathcal{T} }, \hdots, \bar{X}_{N}^{ \mathcal{T} } \} = \{ \bar{X}_{ \mathcal{N} }^{ 1 }, \bar{X}_{ \mathcal{N} }^{ 2 }, \hdots, \bar{X}_{ \mathcal{N} }^{ T } \}$ where 
$ \bar{X}_{i}^{ \mathcal{T} } = \{ X_{i}^{1}, X_{i}^{2}, \hdots, X_{i}^{T}\}$ is time-cumulative subset of $X_{i}$ for time index
 $\mathcal{T}$, and 
 $ \bar{X}_{ \mathcal{N} }^{ t } = \{ X_{1}^{t}, X_{2}^{t}, \hdots, X_{ N }^{t} \}$ 
is a set for the variable set $\mathcal{N}$ given $t$.
Given the set of random variables $\bar{X}_{ \mathcal{N} }^{ \mathcal{T} }$, entropy rate , free entropy, multivariate transfer entropy and residual entropy are defined as follows.
The cumulative sum of entropy rates of variable $i$ at time $T$ is defined as
\begin{eqnarray}
H_{i}^{T} \equiv  \sum_{t = 1}^{ T } H \left( X_{i}^{t} | \bar{ X }_{ i }^{ \bar{t} \setminus t }  \right)
\label{eq-EntropyRateN}
\end{eqnarray}
where $\bar{t} = \{1, 2, \hdots, t\}$ is the cumulative set of time indices 
and $\bar{t} \setminus t = \{1, 2, \hdots, t-1\}$ means set subtraction of index $t$ from $\bar{t}$
with the set subtract operator ``$\setminus$''.
This is identical to the entropy rate defined in a bivariate system \cite{Marko1973,Schreiber2000}.
Free entropy of variable $i$ at time $T$ in the $N$-variable system, which is uncertainty of $x_{i}$ given all the past states of $N$ variables $X^{ \mathcal{T} \setminus T }_{\mathcal{N}}$, is defined as follows.
\begin{eqnarray}
F^{T}_{i} \equiv \sum_{t=1}^{T} H \left( X^{t}_{i}| X^{ \mathcal{T} \setminus T }_{\mathcal{N}} \right)
%= H \left( \bar{X}_{i}^{ \mathcal{T} }, {X}_{ \mathcal{N} \setminus i}^{ \mathcal{T} \setminus T }  \right) 
%- H \left(  X^{ \mathcal{T} \setminus T }_{\mathcal{N}} \right)
\label{eq-FreeEntropyN}
\end{eqnarray}
Multivariate transfer entropy from variable $j$ to $i$ given the set of the other variables $ \mathcal{N} \setminus \{ i, j \}$
in the $N$-variable system $X^{ \mathcal{T} }_{\mathcal{N}}$ is 
defined as follows.
\begin{eqnarray}
 T^{T}_{ j \rightarrow i | \mathcal{N} \setminus \{ i, j \} } 
\equiv
%  \sum_{t = 1}^{T} \sum_{ s = 1 }^{t-1}  I \left(x^{t}_{i};x^{s}_{j}|\bar{X}^{ \bar{t} \setminus t }_{i}, \bar{X}^{ \bar{s} \setminus s }_{j}, X^{ \bar{ t } \setminus t }_{\mathcal{N} \setminus \{ i, j \}} \right)
%= 
\sum_{t=1}^{T} I\left( X_{i}^{t}; \bar{X}_{j}^{ \bar{t} \setminus t } | \bar{X}_{ \mathcal{N} \setminus j }^{ \bar{t} \setminus t } \right)
  \label{eq-MultivariateTransferEntropy}
\end{eqnarray}
Residual information from variable $i$ in the $N$-variable system $X^{ \mathcal{T} }_{\mathcal{N}}$ is defined as follows.
\begin{eqnarray}
R_{ i, j }^{T}
 \equiv
\sum_{t=1}^{T} I\left( X_{i}^{t}; X_{j}^{t} |  \bar{X}_{ \mathcal{N} \setminus \{ i, j \} }^{t} \right)
%= 
%\sum_{k=0}^{N-2}(-1)^{k}
%\sum_{ \mathcal{S} \subset S(\mathcal{N} \setminus \{ i, j \}, k )}
%R_{ i, j | \mathcal{S} }^{T} 
\label{eq-ResidualEntropyN}
\end{eqnarray}
%where 
%\begin{eqnarray}
%$R_{ i, j | s } 
%= \sum_{t = 1}^{T} %\Big\{ %\sum_{k = 2}^{N}
%I\left( 
%x_{i}^{t}; x_{j}^{t}; \bar{X}_{s}^{t}| X_{ \{ i, j \} }^{ \bar{t} \setminus t }, X_{ s }^{ \bar{t} \setminus t }  
%X_{ \mathcal{N} }^{ \bar{t} \setminus t }  
%\right)
%$.
%\end{eqnarray}
Obviously, each of entropies and informations are non-negative: $H_{i} \ge 0$, $F_{i} \ge 0$, $T_{j \rightarrow i | \mathcal{N} \setminus \{i, j\}} \ge 0$, and $R_{ i, j } \ge 0$ for arbitrary $ i \in \mathcal{N}$ and $ j \in \mathcal{ N } \setminus i$. In a special case with a bivariate system ($N=2$), it agrees with the bidirectional information network \cite{Marko1973}.

% \begin{figure}[tb, clip]
%      \includegraphics[width=1\linewidth, clip]{figure/LocalInformationFlows.eps}
%   \caption{\label{fig-LocalInformationFlow} Local in-coming and out-going information flow at variable $i$ in a $N$-variable system.}   %\end{center}
%  \end{figure} 
\subsection{Properties of multivariate information network}
The multivariate information network has the two major properties --holistic and local-- stated in the two theorems.
The first theorem states that, as a whole, a multivariate information network can be viewed as decomposition of total correlation among $N$ variables \cite{Watanabe1960,Garner1962,Studeny1999}.
The second theorem states that, at any node in the network, it holds Kirchhoff's current law or equivalence of the sum of in-coming information flow to the sum of out-going information flows (Figure \ref{fig-MStransfer_entropy}d). 
In addition, each of information flows in the network is always non-negative,
and this allows us to interpret each of informational quantities, entropy rate, free entropy, transfer entropy, and residual entropy, as an amount of information flow. 
The non-negativity of the flows is not trivial under simultaneous satisfaction of the Theorem 1 and 2 below.
The present paper proves that MTE defined in Equations (\ref{eq-EntropyRateN}-\ref{eq-ResidualEntropyN}) has the theoretical
properties stated in the following theorems\footnote{However, note that these theorems may not hold for an empirical MTE estimated
with approximation (e.g., supposed Markov chain and/or stationary of time series) when a dataset violates the assumed approximations.} 
(see also Supporting Information for the details).
%(Figure \ref{fig-LocalInformationFlow}). 
%See also Supplemental Information for the mathematical proofs.
%These wholistic and local properties of information network are formally stated in the following two theorems.

%%%%%%%%%%%%%%%%%%%%%%%%%%%%%%%%%%%%%%%%%%%%%%%%%%%%%%%%%%%%%%%
%% TE Theorem 1: Conditional-TE-decomposition of total correlation
%%%%%%%%%%%%%%%%%%%%%%%%%%%%%%%%%%%%%%%%%%%%%%%%%%%%%%%%%%%%%%%
\begin{thm}[Decomposition of total correlation]
In an $N$-variable system, total correlation 
consists of the sum of all the multivariate transfer entropies and residual entropies.
%$N$-varaible total correlation (Equation \ref{eq-TotalCorrelationNVariables}) can be rewritten as follows.
\begin{eqnarray}
 C( X^{T}_{\mathcal{N}} ) 
=
\sum_{ i = 1 }^{N}\sum_{ j = 1 }^{ i - 1 }
G_{ij}
%\left( 
%T_{ i \rightarrow j | \mathcal{ N } \setminus \{ i, j \} } + T_{ j \rightarrow i | \mathcal{ N } \setminus \{ i, j \} } 
%+
%R_{ i, j } 
%\right)
\label{eq-TotalCorrelationDecompositionNVariables}
\end{eqnarray}
where $
%\begin{equation}
 C( X_{T}^{\mathcal{N}} ) = \sum_{i=1}^{N}H(X_{i}^{T}) - H( X_{T}^{\mathcal{N}} )
%  \label{eq-TotalCorrelationNVariables}
%\end{equation}
$ is total correlation, and 
$G_{ij} =  T_{ i \rightarrow j | \mathcal{ N } \setminus \{ i, j \} } 
+ T_{ j \rightarrow i | \mathcal{ N } \setminus \{ i, j \} } 
+ R_{i,j}^{T}
$ is the sum of transfer entropies and residual entropies.
\end{thm}
%%%%%%%%%%%%%%%%%%%%%%%%%%%%%%%%%%%%%%%%%%%%%%%%%%%%%%%%%%%%%%%
%% TE Theorem 2: Local relationship of multivariate transfer entropy
%%%%%%%%%%%%%%%%%%%%%%%%%%%%%%%%%%%%%%%%%%%%%%%%%%%%%%%%%%%%%%%
%The second theorem states on two identities among free entropy, entropy rate, residual information, and multivariate transfer entropy which guarantees equivalence between the sum of in-coming and out-going information flows at an arbitrary edge in the $N$-variable information network (Figure \ref{fig-LocalInformationFlow}).
\begin{thm}[Local information flow]
Entropy rate of variable $i$ consists of free entropy of variable $i$ and the sum of in-coming multivariate transfer entropies to variable $i$ as follows.
\label{thm-LocalInformationFlow}
\begin{eqnarray}
H_{i}
=
 F_{i} + \sum_{ j \subset \mathcal{N} \setminus i }
 T_{ j \rightarrow i | \mathcal{ N } \setminus \{ i, j \} } 
\label{eq-InComingEntropyRate}
\end{eqnarray}

Entropy rate of variable $i$ can be locally decomposed with the sum of all the in-coming and out-going multivariate transfer entropies and residual entropies.
\begin{eqnarray}
H_{i}
=
 H\left( \bar{X}_{i}^{ \mathcal{T} } | \bar{X}_{ \mathcal{N} \setminus i }^{ \mathcal{T} } \right)
 + \sum_{ j \subset \mathcal{N} \setminus i }
G_{ij}
 \label{eq-OutGoingTransferEntropy}
\end{eqnarray}
%Thus, free entropy of variable $i$ and multivariate transfer entropy from $i$ has the following relationship.
%\begin{eqnarray}
%F_{i}
%= H\left( \bar{X}_{i}^{\mathcal{T}} | \bar{X}_{ \mathcal{N} \setminus i }^{\mathcal{T}} \right)
%+ \sum_{ j \subset \mathcal{N} \setminus i }
%\left( 
% T_{ i \rightarrow j | \mathcal{ N } \setminus \{ i, j \} } + R_{ i, j } 
%\right)
%\label{eq-OutGoingFreeEntropy}
%\end{eqnarray}
\end{thm}

\section{Numerical and empirical studies}
As numerical and empirical validation of the MTE, we report simulation studies with the two classes of nonlinear dynamical systems and two case studies of empirical data analyses.
 Nonlinear dynamical systems would be one of interesting testbeds to demonstrate characterization of directed dependency structure of an unknown system with MTE. In a narrow sense, a generating process of a dynamical system is deterministic. Despite its deterministicity, its long-term chaotic behavior may be unpredictable, and it can be treated as a pseudo random series with knowledge of the initial state at finite precision.
Yet it holds local dependency among variables at each time step. In each of the simulations, we generated a suffficiently long time series from a nonlinear dynamical system with a specific set of parameters. Then we tested whether we can recover the intrinsic dependent relationship on the basis of MTE or PTE applied to the generated time series. %Specifically, we applied the MTE to the Lorenz system and the coupled map lattice.

In the empirical studies, we applied the MTE to two empirical datasets. One is a physiological dataset which has been analyzed in multiple previous studies in the context of a nonlinear time series analysis. The other is a dataset of human body movements in complex actions. 
%Common in these two datasets and other empirical multivariate time series observed in living systems, these complex systems need to coordinate multiple subcomponents in intermediate states neither perfectly static, periodic, nor chaotic. 
Common in these two multivariate time series, complex systems in general need to coordinate multiple subcomponents in order to hold some intermediate states neither perfectly static, periodic, nor chaotic.
Thus it is of great interest to analyze its mutual relationship among multiple subcomponents. % in order to characterize such a complex system.

\subsection{Simulation 1: Lorenz system}
In Simulation 1, we applied the MTE to the Lorenz attractor which is one of well-known three dimensional dynamical systems defined with the following set of ordinary differential equations \cite{Lorenz1963}.
\begin{eqnarray}
 \left\{
\begin{array}{lll}
\frac{ \mathrm{d} x}{\mathrm{d} t} &=& \sigma (y - x)\\
\frac{ \mathrm{d} y}{\mathrm{d} t} &=& x (\rho - z) - y\\
\frac{ \mathrm{d} z}{\mathrm{d} t} &=& xy - \beta z \label{eq-LorenzSystem}
\end{array}
\right.
\end{eqnarray}
where $x$, $y$, and $z$ is the system state, $t$ is time, $\sigma$, $\rho$, and $\beta$  are the system parameters.
Given the ordinary differential equation, we suppose that each differential equation reflects information flows from the past states ($x$, $y$, and $z$) at $t$ to the next states with a short time lag ($t + \Delta t$).
Then the question here is whether we can infer these dependencies by applying information theoretic measures to the generated time series without prior knowledge of the differential equations. In the Lorentz equations, the differential $\frac{\mathrm{d}y}{\mathrm{d}t}$ and $\frac{\mathrm{d}z}{\mathrm{d}t}$ depends on all three variables, meanwhile the differential $\frac{\mathrm{d}x}{\mathrm{d}t}$ depends on $x$ and $y$ but not $z$.
This asymmetric relationship -- the variable $z$ depends on $x$ but not vice versa -- gives a challenge to the measure of dependency in the multivariate system. Based on a good measure of dependency, we can reject the (conditional) dependence from $x$ to $z$ and detect the others.

\begin{figure}[tb, clip]
   \begin{center}
      \includegraphics[width=1\linewidth, clip]{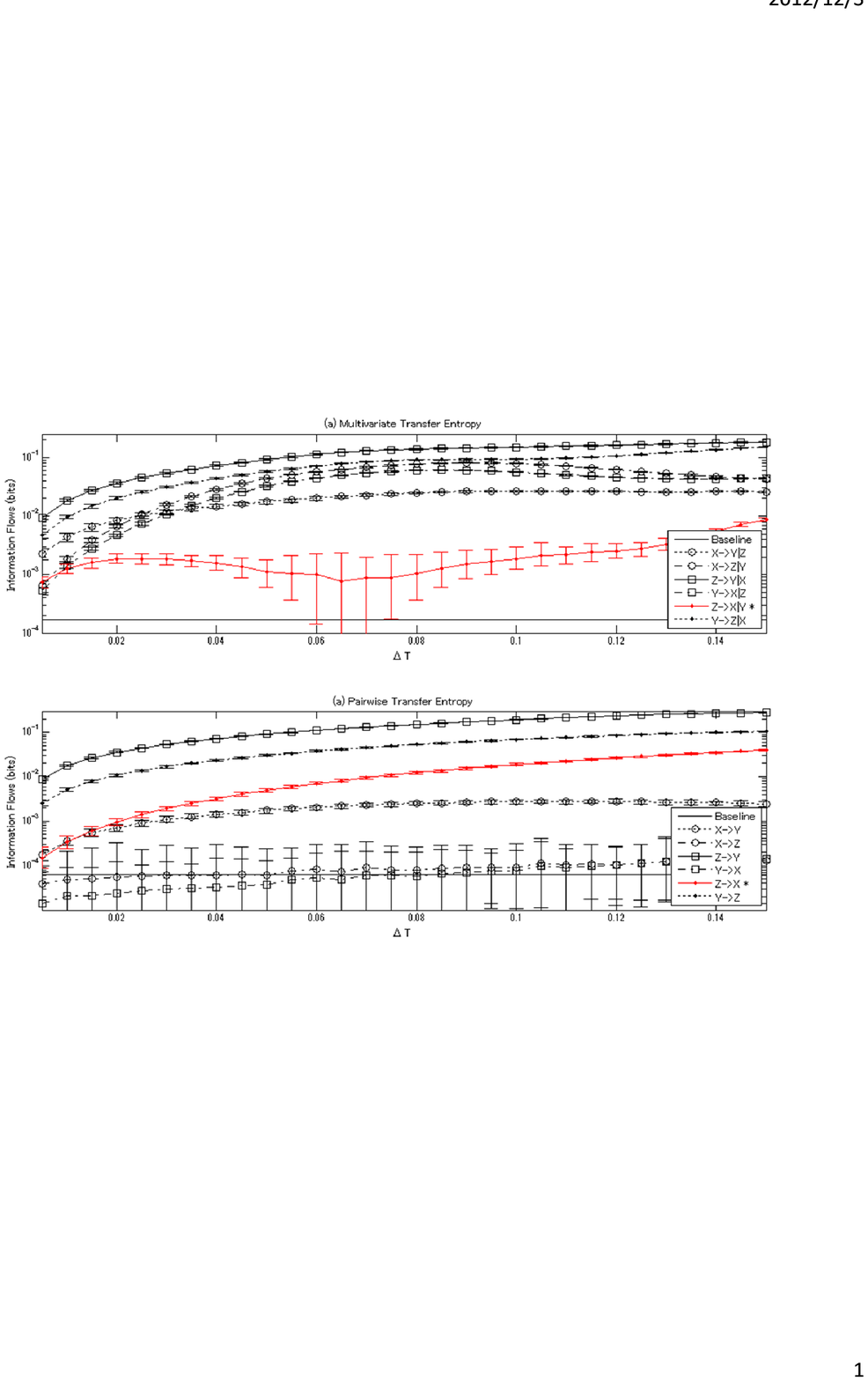}
    \caption{The estimated PTE and MTE for Lorenz attractor as a function of time lag $\Delta t$. The directed pair from $z$ to $x$ to be zero in theory is highlighted in red.\label{fig-SimulationLorenz}} 
  \end{center}
  \end{figure}

Due to the Lorenz system being defined over continuous time, its time series was analyzed by manipulating the time lag $\Delta t$
of the first order Markov chain $\{ t, t + \Delta t \}$ systematically from 0.001 to 0.15.
%The average MTEs and PTEs estimated on the Lorenz system are shown in Figure \ref{fig-SimulationLorenz}.
The upper panel and bottom panel in Figure \ref{fig-SimulationLorenz} show the averages of MTEs and PTEs respectively as a function of the lag $\Delta t$. We performed statistical inference by taking the zero MTE or PTE (conditional independence) as the null hypothesis (See also Method for details), and the upper confidential bounds of the theoretical zero MTE and PTE are shown as a solid line.

The results showed that the MTE from variable $z$ to $x$ (highlighted in red), which is to be zero in theory, was evaluated as the lowest among the six directed pairs at all the lags except for $\Delta t < 0.01$ (Figure \ref{fig-SimulationLorenz}a). At the lags $0.06 < \Delta t < 0.07$, the to-be-zero MTEs from $z$ to $x$ were around the upper confidential bound of the theoretical zero MTE. Meanwhile, the MTE between the other five directed pairs were significant positive at any lags. The results suggested that we could estimate the latent dependent relationship in the Lorenz system on basis of the MTE except for too short lag.

On the other hand, the results showed that the PTE from variable $z$ to $x$ was evaluated as the middle among the six directed pairs of them (Figure \ref{fig-SimulationLorenz}b), and it was significantly larger than the theoretical zero PTE at all the lags $0.001 < \Delta t< 0.15$. One the other hand, the PTE from $y$ to $x$ and from $x$ to $y$, which should be positive in theory, tended to be as low as the theoretical zero PTE at all the lags. These results suggest that PTE does not just overestimate to-be-zero information follow but also underestimate to-be-positive information flows.
In sum, these simulations suggest that the MTE may measure multivariate dependent structure with more accuracy than the PTE. The simulation clearly demonstrated the potential limitation of the PTE when applied to a system with the three variables or more.

\subsection{Simulation 2:  Characterization of various dependent structures}
Simulation 1 suggests the advantage of MTE and potential limitation of PTE in analysis of multivariate systems. In Simulation 2, we analyzed the robustness of MTE-based inference as a function of the number of variables, various topological types of dependent networks, and effects of unobserved noisy variables. Specifically we studied a class of {\it coupled map lattices} (CML) which allows us to systematically manipulate its parameters. The CML is a class of nonlinear dynamical system which linearly combines multiple one-dimensional chaotic systems as subcomponents \cite{Kaneko1992}. 
Although each subsystem behavior is relatively simple and well known, it also shows a global emergent pattern across subsystems
with a particular network topology among subsystems.
Due to these properties, %its simplicity of individual subcomponents and flexibility of the network topology, 
%some variants of CMLs have been used to model various kinds of real-world phenomena  that temporal evolution of each subcomponent follows the same map while we can parametrically manipulate its dependency among subcomponents. 
some variants of coupled map lattice have been used for modeling various kinds of real-world phenomena such as earthquakes \cite{Ceva1995}, form of neurons \cite{Sakaguchi1999}, traffic flow \cite{Yukawa1995}, open flow \cite{Willeboordse1995}, convection \cite{Yanagita1993}, cell-gene interaction \cite{Bignone1993}, epileptic seizures \cite{Larter1999} and so on.
Utilizing its controllability, we analyzed the robustness of MTE based inferences on the system dependency.

\begin{figure}[tb, clip]
   \begin{center}
    \includegraphics[width=1\linewidth, clip]{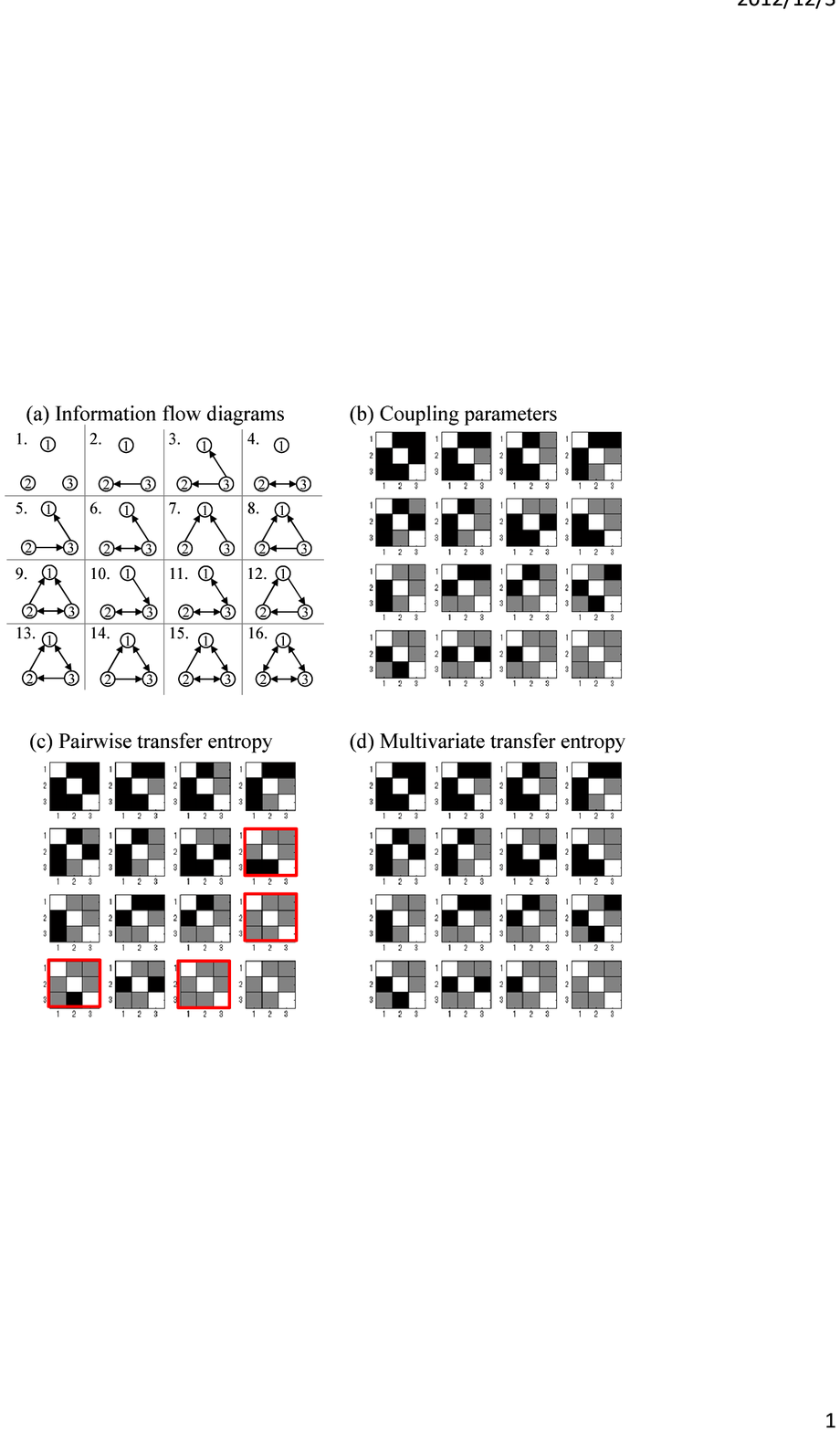}
    \caption{(a) 16 possible diagrams of information flows (its direction indicated by the arrows) among the three variable 1, 2, and 3. (b) The matrices of coupling parameters in the coupled tent map lattice corresponding to the 16 information flow diagrams (white diagonal cell = 1, gray = $\epsilon$, and black = 0), (c-d), The matrices of estimated information flows in which the significant positive PTE and MTE are in gray, the ones not significant are in black, the ones not tested (diagonal cells) are in white. The red outlines of matrices highlight cases with misdetection of information flows. \label{fig-CoupledMapLatticeAll}}       
   \end{center}
 \end{figure}

Specifically, we studied a coupled tent map lattice (CTML) which is defined as follows.
For $0 < x_{i}^{t} < 1$ ($i = 1, 2, \hdots, N$ and $t=0, 1, \hdots, T $), 
\begin{equation}
 x_{i}^{t+1} = f\left( \frac{x_{i}^{t} + \epsilon \sum_{j \neq i} \delta_{ij} x_{j}^{t} + \eta_{i}^t }{1 + \epsilon \sum_{ j \neq i }\delta_{ij} + \eta_{i}^t } \right)
\label{eq-CoupledTentMapLattice}
\end{equation}
where $\epsilon \ge 0$ is the coupling parameter indicating the degree of dependency in the system, $\delta_{ij}$ is either one or zero indicating existence of dependency from $j$ to $i$, $\eta_i^t$ is noise, a random value drawn from an uniform distribution, and $f(x)$ is the so-called tent map in which $f(x) = 2 x$ if $x < \frac{1}{2}$, otherwise $f(x) = 2 - 2 x$.
As in Simulation 1, we define a positive coupling parameter between the variable $x_i^t$ and $x_j^{t+1}$ as positive information flow from variable $i$ to variable $j$. Given binary information flows, either positive or zero, there are 16 different types of dependent networks with directed edges by identifying symmetric topology under exchange of the three variables (Figure \ref{fig-CoupledMapLatticeAll}a).
Each of the 16 diagrams corresponds to a $3 \times 3$ matrix of network topology $\delta_{ij} $ ($i=1,2,3, j=1,2,3$) in Figure \ref{fig-CoupledMapLatticeAll}b. The colored $(i,j)$-cells in the matrices indicate the coupling parameters from $x_j^t$ to $x_i^{t+1}$: white = 1, gray = $\epsilon$, and black = 0 in Equation (\ref{eq-CoupledTentMapLattice}). Given the CTML, we systematically explored all the possible dependency diagrams with three, four, and five variables. There are $2^6$, $2^{12}$, and $2^{20}$ possible combinations of inferences on binary information flows for each dataset, 96, 2616, and 192160 directed pairs in 16, 218, and 9608 unique diagrams for three, four and five variables respectively (Table 1).

Given the latent positive or zero information flows of the CTML, we generated the time series and computed MTE (PTE) for each directed pair. Then we defined MTEs (PTEs) larger than its 99\%-confident upper bound of the theoretical zero MTEs (PTEs) as significant positive information flows. We analyzed the correspondence between the estimated and latent information flows as correct inference. Figures \ref{fig-CoupledMapLatticeAll}c and \ref{fig-CoupledMapLatticeAll}d show the results of estimated dependent pairs in the CTML with the three variables and the coupling parameter $\epsilon = 0.2$ without noise ($\eta_i^t = 0$ for any $t$). The estimated significantly positive MTEs or PTEs are shown as gray, otherwise black (See also Method for the statistical inference). The results showed that the MTE successfully gave correct inferences of dependency for all the 96 directed pairs in the 16 diagrams (Figure \ref{fig-CoupledMapLatticeAll}d). Meanwhile, the PTE overestimated the six directed pairs in four diagrams, and caused incorrect inferences which are highlighted in red in Figure \ref{fig-CoupledMapLatticeAll}c.

All the results of Simulation 2 including the CTML four and five variables are summarized in Table 1. 
The case-based correctness was defined as correct inferences for all the directed pairs in each combination of the diagrams.
For the four-variable CTML, the proportion of correct inference based on the MTE was 90.78\% of the 218 cases and 97.78\% of the 2616 directed pairs. Meanwhile, that based on the PTE was 9.22\% of the cases and 79.67\% of the directed pairs. For the five-variable CTML, the proportion of correct inference based on the MTE was 81.48\% of the 9602 cases and 95.41\% of the 192160 directed pairs. That based on the PTE was 1.24\% of the cases and 71.23\% of the directed pairs.  In sum, the simulation with the CTMLs demonstrated that robustness of MTE across the varying number of variables. On the other hand, inference based on the PTE tended to be more inaccurate as a function of the number of variables.

\subsubsection{Robustness to unobserved variables}
In the next analysis, we tested the robustness of the MTEs when applied to a dataset with unobserved variables. One of lessons derived from Simulation 1 is that PTE for a system with three or more variables may be inaccurate. The same lesson could apply to $N$-variable MTE for ones with $(N+1)$ or more potential variables. Since, unlike the simulations, we cannot always observe all the sufficient set of variables in typical empirical data analyses, it raises a potential concern that MTE may not be better than PTE for the dataset with potential unobserved set of variables. Therefore, it is of practical importance to evaluate robustness of the measures for such datasets with unobserved variables. In the simulation, we generated the time series following the equation $X$ with the noise term $\eta_i^t$ is a random value drawn from the uniform distribution $[0, 0.1]$ for each time step $t$. The random inputs $\eta_i^t$ to the variable $x_{i}^{t+1}$ reflects purtabation by the unobserved set of variables. With or without noisy unobserved variables, we analyzed the average classification performance of the MTE and PTE for the 16 cases of 3-variable CTMLs (Figure \ref{fig-CoupledMapLatticeAll}a) as a function of the coupling rate $\epsilon$ from 0 to 0.6 (Figure \ref{fig-CoupledMapLatticeAll}a and \ref{fig-CoupledMapLatticeAll}b).

\begin{figure}[p, clip]
   \begin{center}
      \includegraphics[width=0.7\linewidth, clip]{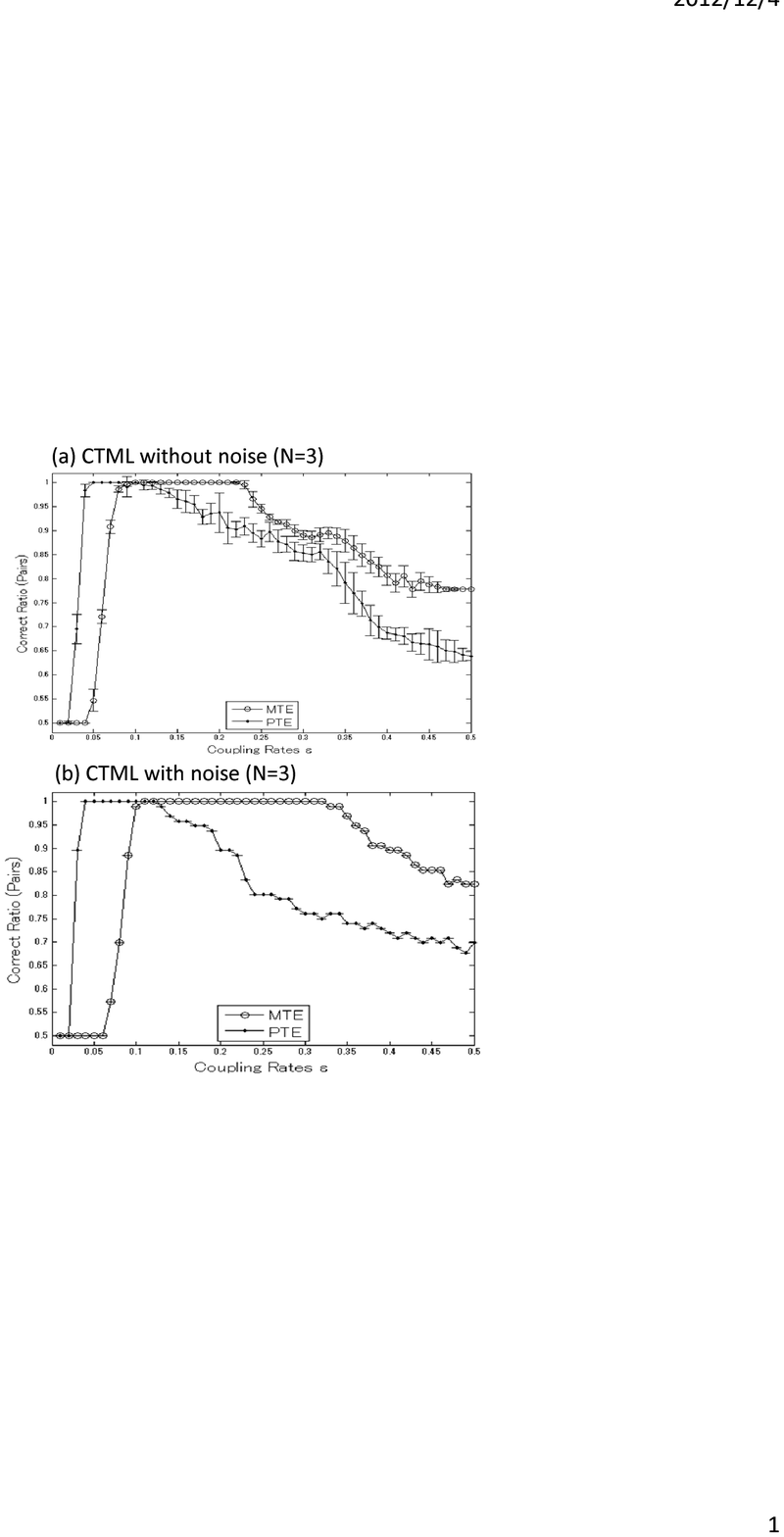}
    \caption{The correct detection of information flows (directed pairs as its unit) as a function of the coupling parameter averaged across the 16 cases of coupled tent map lattice with three variables including (a) no noise factor ($\eta_{i}^t=0$) and (b) noise factors ($0 \le \eta_{i}^t\le 0.1$).\label{fig-CoupledMapLatticeNoise}}   
\end{center}
  \end{figure}

In analysis of the dataset without unobserved variables, on the basis of TEs, we made correct inferences in all the cases at a small range of coupling parameters $0.05 \le \epsilon \le 0.1$ (Figure \ref{fig-CoupledMapLatticeNoise}a). Likewise, on basis of MTEs, we made correct inference in all the cases at a relatively broader range of coupling parameters $0.1\le \epsilon \le 0.25$ (Figure \ref{fig-CoupledMapLatticeNoise}b). Since too large coupling parameters ($\epsilon > 0.25$) made multiple variables perfectly coupled ($R > 0.95$), those coupling variables were difficult to discriminate on the basis of MTEs with its coarse-grain encoding of the phase space. In fact, we found at least one false detection due to nearly perfect coupling in Cases 10, 11, 13, 14 and 15 (Figure \ref{fig-CoupledMapLatticeAll}a). In contrast, the advantage of PTEs in a small coupling parameter is likely to be caused by the relatively small effects of the third variable. In addition, MTE needs to estimate the probabilistic distribution a large combinatorial space relative to the given sample size. This sparseness of samples leads MTE to be more conservative to reject false positive information flows. Except for this small coupling rate advantage, MTE outperformed PTE in most of the cases and parameters.

 Simulation of the datasets with noisy latent variable showed basically similar patterns as found in that of the noiseless dataset. In Figure \ref{fig-CoupledMapLatticeNoise}b, we found the advantage of PTE in a small coupling parameter $\epsilon = 0.1$ or smaller and the better detection performance of MTE otherwise. We also found a remarkable difference from the noiseless dataset: MTE tended to show even better performance while TE tended to show worse performance in the dataset with noisy latent variable. In this particular simulation, the reason for the even better performance of MTE for the noisy data was perhaps because the noisy variables decoupled the perfect-coupled variables (at $\epsilon > 0.25$) which harmed MTE performance for the noiseless dataset. In sum, these results suggested the robust relative advantage of MTE even with noisy latent variables.

\subsubsection{Summary of numerical studies}
We summarize the findings in Simulations 1 and 2 in the four points. First, MTE showed advantages over PTE in both continuous-time system asymmetric under exchange of variables (Simulation 1) and discrete-time systems (Simulation 2). Second MTE also showed advantages in various types of dependent topology in the CTMLs with 3, 4, and 5 variables. Third, its advantage is robust even in the analysis of the datasets with unobserved noisy variables. Finally, the simulations also showed a limitation of the MTE, more conservative estimate of information flows than PTE. We will discuss this technical limitation in the later section.

\subsection{Empirical data analysis 1: Physiological data}
As a case study, we analyzed a physiological dataset including three vital signs recorded in a sleeping person\cite{Rigney1993,Ichimaru1999}. Besides being a trivariate time series, we chose this dataset as a benchmark test, since it has been analyzed across many theoretical studies \cite{Kantz1997,Schreiber2000,Angelini2007,Marinazzo2008}. The original data consists of the set of three time series of heart rates, breath rates, and blood oxygen concentration recorded at 0.5 Hz of sampling rate. The particular person measured has been known to show respiratory sinus arrhythmia. It is a frequently-seen symptom that shows correlation between heart rates and breath rates. As expected, the previous study showed that the heart rates and breath rates transferred information bidirectionally by applying PTE \cite{Schreiber2000}. However, as suggested in Simulation 1 and 2, it is potentially possible to have such seeming information transfers caused by the third factor, for example, the blood oxygen concentration in this dataset. Thus, we performed reanalysis on the dataset not just as bivariate but as a part of a trivariate system by applying the MTE.

\begin{figure}[p, clip]
   \begin{center}
      \includegraphics[width=1\linewidth, clip]{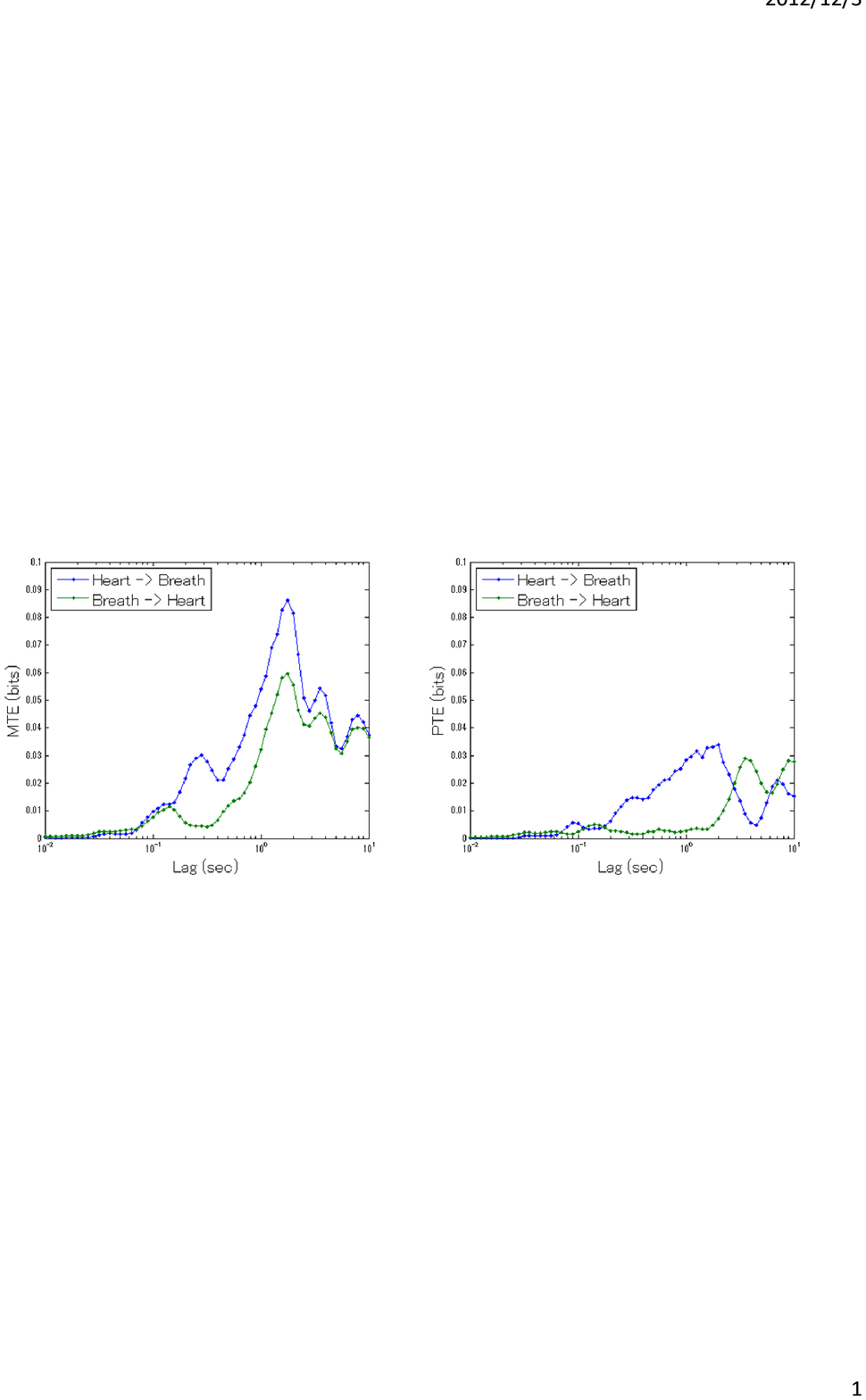}
    \caption{The estimated information transfer between heart rates and breath rates with MTE given blood oxygen concentration (left) and PTE (right).\label{fig-SantaFeB}}   \end{center}
  \end{figure}

Figures \ref{fig-SantaFeB}a and \ref{fig-SantaFeB}b respectively show the PTE (as bivariate series) and MTE (given the blood oxygen concentration) between heart rates and breath rates as a function of time lag. In Figure \ref{fig-SantaFeB}b, we replicated the qualitative patterns of PTEs as found in the previous study: both directions have information transfers at most of time lags, while the heart rate tended to transfer information to the breath rates more than the other direction \footnote{It is potentially possible to have the results disagreed in the present and previous study due to its technical difference in estimation method and choice of a particular subset of time series.}.
%Thus, it is necessary to confirm its qualitative agreement between different estimation techniques.

The qualitative patterns of PTE and MTE basically agreed - heart rates and breath rates are tightly coupled bidirectionally with or without respect to blood oxygen concentration. This result confirmed the conclusion in the previous study even as a part of a trivariate system in regard to these qualitative patterns. However, we also found a difference between the two measures. In MTE, we also found that information transfer between the heart rates and the breath rates peaked around the same time scale of the lag approximately 2 sec. One the other hand, in PTE, the two directions had peaks at quite different scales of time lags: PTE from Heart to Breath peaked at approximately 2 sec and that from Breath to Heart was at approximately 20 sec. At this moment, we could not conclude which of the results, synchronized or delayed peaks in MTE and PTE, is more plausible in light of empirical findings. It is an open question for further empirical studies.

\subsection{Empirical data analysis 2: Motor coordination in complex actions}
The second case study is an analysis of complex human actions.
Our bodily actions require coordinated movements of multiple body parts. A human body consists of over two hundreds bones, numerous muscles, and billions of neurons in central and peripheral nerve systems controlling them with feedback loops. Obviously, making a smooth action requires integrated control over all levels of these systems. It is of our interest to characterize human motor coordination in skillful actions through the MTE. Specifically, we chose a dataset of complex actions performed by multiple players with different expertise levels. The data was originally obtained in order to analyze the levels of expertise in the samba music plays \cite{Yamamoto2006,Yamamoto2008}. %(Fujinami & Yamamoto). 
The original dataset consists of five players, and each player performed basic samba shaking actions in five different tempos (60, 75, 90, 105, 120 beats per minute, and each trial lasted 97.4 seconds on average) by being cued with a metronome. While playing, three dimensional motions of 18 markers, attached on body parts and musical instruments, were recorded at 86.1Hz of sampling rate.

As well as the original study, here we aim to find the relationship between informational properties among bodily actions and the expertise levels in the motor skill. The present study analyzed the actions of three players, chosen from the original five, who are one master player (more than thirty years of experience) and two of his disciples Disciple 1 (six years of experience) and Disciple 2 (two years of experience). The expertise levels between the master and his disciples were expected to be different. Given our knowledge of the players' expertise levels as the ground truth, we tested whether MTE can successfully detect the differences in their skill levels. For simplicity, we limited ourselves to analyze a subset of the original datasets, 3190 samples (74.1 seconds long) of four motions of markers attached on right wrist, right elbow, and two sides of the musical instrument (shaker). These were the essential parts of the samba actions making sounds directly, and we expected that information flows among them would be crucial to characterize the players' expertise levels. In a smooth samba play, multiple body parts need to be coordinated to perform the complex actions. Thus, perhaps common in general multivariate time series analyses, one of challenges in this analysis is to decompose the smooth actions into information flows between body parts.

\begin{figure}[p, clip]
   \begin{center}
      \includegraphics[width=0.6\linewidth, clip]{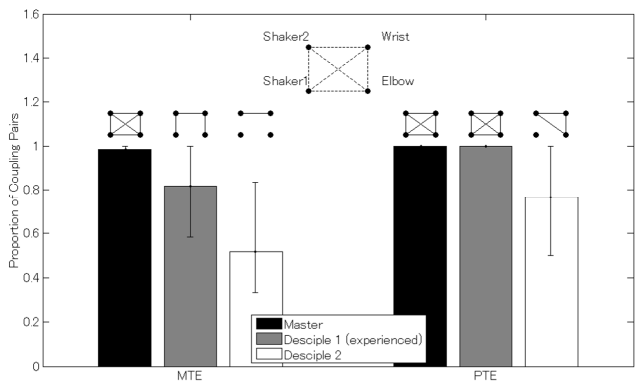}
    \caption{the proportion of coupling bodily movements estimated with MTE and PTE. The proportions for each subject are average across 5 different tempo conditions, and its maximum and minimum are shown as the errorbar. Each graph on top of the bar shows informational connectivity among four bodily movements in which an edge shows one of either or both directions have significant MTEs or PTEs across all the five conditions.\label{fig-Samba}}   \end{center}
  \end{figure}

In the analysis, we applied the MTE and PTE to all the directed pairs of four motions. Figure \ref{fig-Samba} shows the proportion of directed pairs with significantly positive MTE and PTE averaged across five different tempo conditions ($60 = 12 \times 5$ directed pairs in total for each subject). The results showed distinguishable patterns of informational coupling among the master and the two disciples. Across all the five tempo conditions, we found all the body parts in the master player nearly perfectly coupled. Meanwhile, Disciple 2 with the least experience among the three showed the least number of informational coupled pairs. In Figure \ref{fig-Samba}, each graph on the top shows the information network of each player. It has a solid edge between the markers if at least one of the two directions had significant positive MTE across all the five tempo conditions. The graph of Disciple 2 shows that the only consistently coupling pair was his wrist and a side of shaker. That of Disciple 1 showed the three edges, elbow-wrist, wrist-shaker2, and shaker2-shaker1, which suggests these physically connected parts formed a action like whip stroking. As expected, these results showed consistency between the player's expertise levels and the MTE-based informational properties in their bodily actions.

As a comparison, we also applied PTE to the same dataset (Figure \ref{fig-Samba}). The results showed that PTE did not detect the differences between the master and Disciple 1 both of whom showed significant PTEs in all the directed pairs. Compared with the MTE, PTE tended to overestimate the coupling pairs in all three players. In the graph patterns, PTE detected positive information between the pairs which MTE did not detect. Regarding our knowledge of subjects' expertise levels, PTE estimation was likely to detect false positive information flows due to the effects of the two other unconsidered variables. As a result, we could not find the difference in the PTEs among players as clearly as found in the MTEs. These results of empirical data analyses suggested potential applicability of MTE to empirical complex multivariate time series.

\section{Discussion}
The present study proposes an extended information theoretic measure for system characterization through time series. The multivariate transfer entropy is a natural generalization of pairwise transfer entropy to a multivariate system holding a set of theoretical properties. The MTE was tested on the two classes of nonlinear dynamical systems and on the two empirical datasets. The simulations demonstrated the advantage of MTE over PTE, in both discrete-time and continuous-time systems, with most of the topologies of dependency among 3, 4, and 5 variables, and even with additional noisy latent variable. 
%These advantages of MTE would stem from the two theoretical properties, decomposition of higher order dependency as information flows in a network, which the PTE does not always satisfy for a system with three or more variables. 
These advantages of MTE would stem from the theoretical property that the MTE decomposes higher order dependencies into information flows in a network. Since the PTE does not always satisfy it for a system with three or more variables, the PTE from $A$ to $B$ may take some value
independently of the PTE from $A$ to $C$. As a result, the PTE from $A$ to $B$ may overestimate or underestimate dependency between $A$ and $B$
when the third variable $C$ also has effects on $B$.

Application to the two empirical datasets suggested its potential use in the analysis of empirical complex systems including physiological signals and human motor coordination. In analysis of such datasets with complex interactions, MTE is likely to be useful because it exclusively measures a pair of variables by cancelling out the effects from the other variables. This general applicability of MTE to multivariate systems covers a broader range of empirical and theoretical fields using PTEs \cite{Lizier2008,Quinn2010,Bauer2007,Nichols2005,Rubinov2010,Vicente2011,Baek2005,Moniz2007,Kulp2009,Hidaka2010,Hidaka2010b,Sumioka2007}.

%\subsubsection{ Decomposition/ reconstruction of information network toward non-parametric structure learning in a graph model}
%Typically, probabilitic variables are represented as nodes in a graph either unidirected or directed.
% Learning structure (Ghahramani, ``Graphical model: Parameter learning'') by Presentation of Ho sensei's 
%either unidirected graphs or 

\subsection{Technical limitations and future works}
In contrast to its relatively robust and accurate evaluations, the present simulation studies also suggested a limitation of the MTE. The analysis in Simulation 2 showed that the MTE was more conservative to detect dependency than the PTE. This problem of MTE was likely to be caused by the technical issue in its estimation. The $N$-variable MTE needs conditional entropy of $N$ variables in which the combinatorial space may grow as an exponential function of the number of variables. It causes high computational costs and inaccuracy of the estimation due to the sparsity of samples relative to the exponentially growing space. In the current implementation, which was not optimized but was computed in a naive way, it was very costly to compute even relatively modest number of variables $N > 5$. This estimation problem prevented us from using a finer-grained binning on the phase space, and we suggest that it resulted in the conservative detection of information by MTE. Therefore, futher work could include developing a technique relaxing this problem. Similar technical issues have been discussed for PTE such as small-sample correlation for pairwise transfer entropy \cite{Marschinski2002} or non-parametric probabilistic density estimator for continuous time series \cite{Kaiser2002}.
%Recently, an efficient algorithm was proposed for a limited class of graphical modeling such as dependency tree in \cite{Quinn2010} which extends \cite{Chow1968} with bivariate directed measure of dependency.
%Thus, combined with an efficient search and threshold (by extending the existing efficient algorithm by \cite{Chow1968,Quinn2010}), we expect a reasonable scale up in practical estimation.

Another related concern in empirical analyses is parameter specification. In order to accurately measure dependency, we need to specify temporal delay and estimators of probabilistic distributions. The current simulations were demonstrated with one of the simplest probabilistic models - binary coding by median splitting with the first order Makov chain. It is an open question to what extent estimation of the MTE depends on the choice of these parameters which perhaps depends on the case. More importantly, how can we choose the probabilistic model and delays? One of potential solutions for this problem for dynamical systems is {\it generating partitions}. A set of generating partitions gives a theoretical ground for "best" discrete states of discrete or continuous dynamical system which has one-to-one correspondence between a series of symbols and a subset of state space. It has been constructed for several low dimensional chaotic systems (e.g., coupled map lattice \cite{Pethel2006} and H\'enon map \cite{Biham1990})
and several algorithms estimating symbolic dynamics from an empirical time series have been proposed \cite{Kennel2003,Buhl2005,Hirata2004}.

\section{Methods}
In all the analyses in the present study, a set of continuous time series of T samples of N variables was converted to a symbolic form encoding the original data in a coarse grained representation (See each analysis below for the specific symbolization process). Based on the symbolic series, probabilistic distribution of time series assuming K-th order Markov chain was estimated, then MTE and PTE were estimated on the probabilistic distribution subsequently (See Estimation of probabilistic distribution below for details). The estimated MTE (PTE) was compared to the corresponding theoretical zero MTE (PTE) (See Zero transfer entropy below). All the computational routines used in the simulations and empirical data analyses were written on the MATLAB plat form, which is available at the author's website (http://www.jaist.ac.jp/\~{}shhidaka/).

\subsection{Estimation of probabilistic distribution}
In each simulation above, 
an array of the values $x_{i, t}$ for $i = 1, 2, \hdots, N$, $t = 1, 2, \hdots, L$ was given as the $N$-dimensional
time series of length $L$. Each value in dimension $i$ was symbolized $\hat{x}_{it} = f( x_{it} )$ where 
$f_{M}(x): R^{1} \rightarrow \mathcal{M}_{i}^{1}$ is the symbolization function mapping the one dimensional real space
to the symbol set of $M_{i}$ alphabets, $\mathcal{M}_{i} = \{ 1, 2, \hdots, M_{i}\}$, which is specified in each simulation. 
The $N$-tuple, $y_{t} = \{ \hat{x}_{ 1, t}, \hat{x}_{ 1, t}, \hdots, \hat{x}_{ N, t} \} \in \mathcal{M}^{N}$, was treated as the joint space of $N$ symbols.
Assuming stationarity and $K$-th order Markovity, 
the conditional probabilistic distribution $P( y^{t} | y^{t-1}, y^{t-2}, \hdots, y^{t-K} )$ over $\{ t-K, t-K+1, \hdots, t\}$ was estimated by its maximum likelihood estimator (MLE) assuming the multinomial distribution over the joint symbol space 
$\{ \mathcal{M}_{1}^{K} \otimes \mathcal{M}_{2}^{K} \otimes \hdots \otimes \mathcal{M}_{i}^{K}\}$. 
Specifically, the MLE is the frequency over
the joint symbol space $\{ \mathcal{M}_{1}^{K} \otimes \mathcal{M}_{2}^{K} \otimes \hdots \otimes \mathcal{M}_{i}^{K}\}$ which is normalized to be probability.
According to the stationary assumption, $P( y^{t_{1}} | y^{t_{1}-1}, y^{t_{1}-2}, \hdots, y^{t_{1}-K} ) = 
P( y^{t_{2}} | y^{t_{2}-1}, y^{t_{2}-2}, \hdots, y^{t_{2}-K} )$ for any $t_{1} \le L-K$ and $t_{2} \le L-K$.
Thus, a series within the time window of $K+1$, $z_{t}^{t+K} = \{y_{t}, y_{t+1}, \hdots, y_{t+K}\}$,
was counted across the data length $t = 1, 2, \hdots, L - K$, thus, a dataset of length $L$ provides $L-K$ samples for estimation of the conditional probabilistic distribution $P( y^{t} | y^{t-1}, y^{t-2}, \hdots, y^{t-K} )$.
Since the joint symbol space has a large number of possible combinations $\prod_{i}^{N} M_{i}^{K}$
growing as exponential function of dimension $N$ and time window length $K$, an empirical dataset with a limite sample size
may be too sparse to estimate probabilistic distribution over the joint symbolic space.
Therefore, another reasonable choice for a sparse-data estimator would be the ones combined with various kinds of smoothing 
techniques on the $K$-gram model. 
The author confirmed that the modified Kneser-Ney smoothing \cite{Chen1999} was effective in particular for data with small sample size,
although the present paper reported the MLE estimator in all the analyses.
%Although the present paper reported the MLE on the sufficient long time series $L \ge 5 \times 10^4$,
% the Kneser-Ney smoothing estimator provided accurate estimation than the MLE for the time series of length shorter than $L \le 10^4$ in the case $M_{i}=2$ ($i=1,2,N$), $N=3$, $K=2$.

\subsection{Zero transfer entropy}
The estimate transfer entropy was compared with the null hypothesis that the true (pairwise or multivariate) transfer entropy is zero meaning conditional independence of a given pair of variables. Note that, even under such the null hypothesis, estimated transfer entropy may be positive due to a finite sample size for estimation. The probabilistic distribution of the null transfer entropy (as a special case of conditional mutual information) follows gamma distribution with the shape parameter $s_1$ and the scale parameter $s_2$ \cite{Goebel2005}.
%(See also \cite{} the related works about Bayesian estimator of the mutual information). 
The parameters $s_1$ and $s2$ may vary across simulations due to specific features of samples, but its maximum is X and Y in Simulation 1 and Z and W in Simulation 2, XX and XX in the empirical data analysis.

\subsection{Simulation 1} 
In Simulation 1, we generated a time series from the Lorenz  system  from $t_0 = 50$ to $t  = 2050$  with  the  initial value $\{ x(t_0), y(t_0), z(t_0) \} = \{ 1 + \eta_1, 1/2 + \eta_2, 0 + \eta_3 \}$ and  the parameters $\{\sigma, \rho, \beta \} = \{ 10, 28, 8/3 \}$ (Equation \ref{eq-LorenzSystem}) where each of the noise factors $\{\eta_1, \eta_2, \eta_3\}$ is a random value drawn from uniform distribution from 0 to 0.01. Using the solver of the ordinary differential equation (ode45 routine in the MATLAB), we obtain approximately 118,000 samples of the three dimensional series, and resample the time series by linear interpolation as desired temporal resolution from 0.001 to 0.15 per sample. For each of given temporal resolutions $\Delta t$, we obtained 300,000 samples by taking a subset of $c$ sets of the time series with different initial values where $c = [\frac{300000 \Delta t}{2000}])$ and $[x]$ is the maximum integer equal or smaller than $x$. Given a set of 3 dimensional time series, in order to form probabilistic distribution, we convert each variable to binary series $s_{i}^t$ of dimension $i$ and time $t$ by splitting median point of each dimension. The first order Markov chain $p(s_{i}, s_{i}^{t+\Delta t})$ (for $i = 1, 2, and 3$) was used for MTE and TE estimation.

\subsection{Simulation 2}
 We generate time series $x_i^t$ of the coupled tent map lattice based on the Equation \ref{eq-CoupledTentMapLattice} with a given set of parameters (coupling parameter $\epsilon$ and $\delta_{ij}$ for $i = 1, 2, \hdots , N, j = 1, 2, \hdots , N$ in Equation \ref{eq-CoupledTentMapLattice} of $N$ variables), a set of random values $\eta_{i}^{t}$ drawn from $(0, r)$ $r$ is either 0 or 0.1, and a set of random initial values $x_1^t$ drawn from uniform distribution of the range $(0, 1)$. In each case, a time series of $10^5$ steps after the first 1000 samples discarded as transient was converted to binary values $s_i^t$ by median splitting $s_t^t = h(x_i^t \ge \bar{x}_i^t)$ where $\bar{x}_i^t$ is the median of $x_{i}^t$ ($i=1,2, \hdots,10^5$) ) and the Heaviside function $h(x)$ is 1 if $x$ is positive 0 otherwise. The third order Markov chain $p(s_{i}, s_{i}^{t+2 }, s_{i}^{t+2 }, s_{i}^{t+4 })$ (for $i = 1, 2, 3$) was used for MTE and TE estimation.

\subsection{Empirical data analysis 1}
We analyzed the dataset B of the trivariate time series, heart rates, breath rates, and blood oxygen concentration, retrieved from the Santa Fe time series competition (http://www-psych.stanford.edu/\~{}andreas/Time-Series/SantaFe.html). We concatenated all the consecutive time series longer than 250 seconds and in the waking states diagnosed by the expert, and made a trivariate time series of 11560 samples. For each of given temporal resolutions $\Delta t$, we obtained 20,000 samples by taking subset of $C$ sets of the time series with different lags $\{ t_0 + t, t_0+t+\Delta t, \hdots, t_0+t+k \Delta t\}$ where $t = 0, \frac{ \Delta t }{c}, \hdots, \frac{ (c-1) \Delta t }{c}$, $c = [\frac{ 20000 \Delta t }{11560} ])$ and $[x]$ is the maximum integer equal or smaller than $x$. 

\subsection{Empirical data analysis 2}
The dataset consists of three players performing in five different tempos (60, 75, 90, 105, 120 beats per minute, and each trial lasted 97.4 seconds on average). The movements of four markers attached on right elbow, right wrist, and the two sides of musical instruments, each originally recorded at 86.1 Hz, were analyzed. In the analysis, after down-sampling the original data to 46.05 Hz, the first 250 samples (5.81 second long from the beginning of the recording) were excluded as initial setup of the actions, and 3250 samples (75.5 second long) of movements were analyzed for each subject. In order to reduce measurement noise, for each movement of the markers, the local linear projective method was performed after phase space reconstruction of each time series on the 31 dimensional time delay space with 46 msec (i.e., $\{ t, t + \Delta t, t+2\Delta t, \hdots, t + 30 \Delta t\}$ where $\Delta t= 46$ msec) \cite{Kantz1997}. For each estimated phase space, a symbol series was assigned using the symbolic false nearest neighbor method \cite{Buhl2005} which estimates a generating partition for a time series. Given the four-variable symbolic series, we applied the multivariate transfer entropy. The estimated MTE greater than the zero MTE at the level of $p<0.001$ were defined as a significant MTE for each of the four body parts in each condition.

%% == end of paper:

%% Optional Materials and Methods Section
%% The Materials and Methods section header will be added automatically.

%% Enter any subheads and the Materials and Methods text below.
%\begin{materials}
% Materials text
%\end{materials}

%% Optional Appendix or Appendices
%% \appendix Appendix text...
%% or, for appendix with title, use square brackets:
%% \appendix[Appendix Title]

\begin{acknowledgments}
The author is grateful to Dr. Tsutomu Fujinami for his kind offering of his dataset of a set of complex human actions.
The author thanks Dr. Neeraj Kashyap, Dr. Brian Kurkoski, Takuma Torii, and Akira Masumi for their helpful discussions and comments on the early version of the manuscript. This work was supported by 
Artificial Intelligence Research Promotion Foundation and Grant-in-Aid for Scientific Research B No. 23300099.
%-- text of acknowledgments here, including grant info --
\end{acknowledgments}

%% PNAS does not support submission of supporting .tex files such as BibTeX.
%% Instead all references must be included in the article .tex document. 
%% If you currently use BibTeX, your bibliography is formed because the 
%% command \verb+\bibliography{}+ brings the <filename>.bbl file into your
%% .tex document. To conform to PNAS requirements, copy the reference listings
%% from your .bbl file and add them to the article .tex file, using the
%% bibliography environment described above.  

%%  Contact pnas@nas.edu if you need assistance with your
%%  bibliography.

% Sample bibliography item in PNAS format:
%% \bibitem{in-text reference} comma-separated author names up to 5,
%% for more than 5 authors use first author last name et al. (year published)
%% article title  {\it Journal Name} volume #: start page-end page.
%% ie,
% \bibitem{Neuhaus} Neuhaus J-M, Sitcher L, Meins F, Jr, Boller T (1991) 
% A short C-terminal sequence is necessary and sufficient for the
% targeting of chitinases to the plant vacuole. 
% {\it Proc Natl Acad Sci USA} 88:10362-10366.

%\nocite{Williams2011}

%% Enter the largest bibliography number in the facing curly brackets
%% following \begin{thebibliography}

%\bibliographystyle{apacite}
\bibliographystyle{pnas}
\bibliography{GTEReferences}

%\begin{thebibliography}{}
%
%\end{thebibliography}

%\section{Figure legends}

%\section{Table legends}

\end{article}

\begin{table}[htbp]
%\centering
\caption{Summary of the inference performance of information flows in the coupled map lattice with 3, 4 and 5 variables and coupling parameter $\epsilon = 0.2$.}%\label{tab-CoupledMapLatticeAll}}
\begin{tabular}%{\hsize}{@{\extracolsep{\fill}}lcr}%
{|c|c|c|c|c|c|c|c}
\hline
%~&~ & ~ & ~ & Multivariate correct & ~ & Bivariate correct & ~ \\ \hline
\multicolumn{4}{|c|}{ Simulation settings }& \multicolumn{2}{|c|}{ Multivariate TE } & \multicolumn{2}{|c|}{ Pariwise TE } \\ \hline\hline
N & Comb. & Cases & Pairs & Cases & Pairs & Cases & \multicolumn{1}{|c|}{Pairs}\\ \hline
3 & $2^{6}$ & 16 & 96 & 100\% & 100\% & 75.0\% & \multicolumn{1}{|c|}{ 93.75\% } \\ \hline
4 & $2^{12}$ & 218 & 2616 & 90.78\% & 97.78\% & 9.22\% & \multicolumn{1}{|c|}{ 70.67\% }\\ \hline
5 & $2^{20}$ & 9608 & 192160 & 81.48\% & 95.41\% & 1.24\% & \multicolumn{1}{|c|}{ 71.23\% }\\
%64, 4096, 1048576
\hline
\end{tabular}
\end{table}

%%%%%%%%%%%%%%%%%%%%%%%%%%%%%%%%%%%%%%%%%%%%%%%%%%%%%%%%%%%%%%%%

%% Adding Figure and Table References
%% Be sure to add figures and tables after \end{article}
%% and before \end{document}

%% For figures, put the caption below the illustration.
%%
%% \begin{figure}
%% \caption{Almost Sharp Front}\label{afoto}
%% \end{figure}

%% For Tables, put caption above table
%%
%% Table caption should start with a capital letter, continue with lower case
%% and not have a period at the end
%% Using @{\vrule height ?? depth ?? width0pt} in the tabular preamble will
%% keep that much space between every line in the table.

%% \begin{table}
%% \caption{Repeat length of longer allele by age of onset class}
%% \begin{tabular}{@{\vrule height 10.5pt depth4pt  width0pt}lrcccc}
%% table text
%% \end{tabular}
%% \end{table}

%% For two column figures and tables, use the following:

%% \begin{figure*}
%% \caption{Almost Sharp Front}\label{afoto}
%% \end{figure*}

%% \begin{table*}
%% \caption{Repeat length of longer allele by age of onset class}
%% \begin{tabular}{ccc}
%% table text
%% \end{tabular}
%% \end{table*}

\end{document}

\maketitle
%\tableofcontents

%\begin{picture}(200,60)(0,0)
%　\put(0,0){\dashbox{2}(80,60){作図領域}}
% \put(0,0){\vector(1,0){80}}
% \put(0,0){\vector(0,1){60}}
%end{picture}
\section{Outline}
This supporting information provides mathematical proofs of the theorems on decomposition of total entropy (Theorem 1) and local information flows (Theorem 2).
The two theorems stand on three lemmas: Chain rule of multivariate conditional mutual information (Lemma 1), Local decomposition of multivariate conditional mutual information (Lemma 2), and decomposition of joint entropy by multivariate mutual information (Lemma 3).
Theorem 1 needs all the three lemmas, and Theorem 2 needs Lemma 3. 
Though all the proofs, it is necessary to assume neither stationary nor Markovity, since it considers the state at each step depends on all the past states from an initial time step to the previous state.
All the three lemmas are the results of elementary derivation from the basic formulation of information theory including entropy, conditional entropy, joint entropy, mutual information, and conditional mutual information, and multivariate mutual information.
Therefore, we begin with confirmation of basic mathematical properties of entropy and its variants. 
Here most of the notations follow convention of terms in information theory in Cover \& Thomas \cite{CoverThomas1991}
 except for the specific terms in the bidirectional information \cite{Marko1973,Schreiber2000}.

\section{Basic formulae in information theory}
Let $X$ be the probabilistic variable over the set of symbol $\mathcal{M} = \{1, 2, \hdots, M\}$ and $p(x)$ be probability
of a drawn symbol $x \subset \mathcal{M}$.
Then the entropy of the probabilistic variable $X$ is as follows.
% and $p( X )$, and $p(x^{(n)})=p(x_{1}, x_{2}, ..., x_{n})$ 
%denotes joint probability of a series of $\{x_{1}, x_{2}, ..., x_{n}\}$.
%{\bf Shannon entropy}
\begin{equation}
 H(X) = -\sum_{x}p(x)\log p(x)
  \label{eq-Entropy}
\end{equation}
where $\sum_{x}$ denotes the summation over all the symbols $x \subset \mathcal{M}$.
Joint entropy for joint probability of $N$ probabilistic variables $\{ X_{1}, X_{2}, \hdots, X_{N}\}$ is as follows.
%{\bf Joint entropy}
\begin{equation}
 H(X_{1}, X_{2}, \hdots, X_{N})= -\sum_{x_{1}, x_{2}, \hdots, x_{N}} p(x_{1}, x_{2}, \hdots, x_{N}) \log p(x_{1}, x_{2}, \hdots, x_{N})
  \label{eq-JointEntropy}
\end{equation}
where $\sum_{x_{1}, x_{2}, \hdots, x_{N}}$ denotes the summation over all the direct product set $\{x_{1}, x_{2}, \hdots, x_{N}\} \subset \mathcal{M}_{1}\otimes \mathcal{M}_{1} \otimes \hdots \otimes \mathcal{M}_{N}$ and $\mathcal{M}_{i} = \{1, 2, \hdots, M_{i}\}$.
Then, conditional entropy of $N$ probabilistic variables $\{ X_{1}, X_{2}, \hdots, X_{N}\}$ given $Y$ is as follows.
\begin{equation}
 H(X_{1}, X_{2}, \hdots, X_{N} | Y )= -\sum_{x_{1}, x_{2}, \hdots, x_{N}, y} p(x_{1}, x_{2}, \hdots, x_{N}, y) 
%\left(
%\log p(x_{1}, x_{2}, \hdots, x_{N}, y )
%- \log p( y )
%\right)%
\log p(x_{1}, x_{2}, \hdots, x_{N} | y )
  \label{eq-ConditionalEntropy}
\end{equation}
where $p(x_{1}, x_{2}, \hdots, x_{N} | y ) = \frac{ p(x_{1}, x_{2}, \hdots, x_{N}, y ) } { p(y) }$ is the conditional probability of the set of probabilistic variables given variable $y$.
Obviously, joint entropy and conditional entropy are both non-negative, as non-negativity of self-information $-\log p(x) \ge 0$
and $-\log p(x | y) \ge 0$. %{\bf (TODO)} XX inequality respectively.
Using Bayes's theorem $p(x|y) = \frac{p(y,x)}{p(y)}$, joint entropy $H(X_{1}, X_{2}, \hdots, X_{n})$ is identical to the sum of conditional entropy.
\begin{equation}
 H(X_{1}, X_{2}, \hdots, X_{N}) = \sum_{i = 1}^{N} H(X_{i}| X_{1}, X_{2}, \hdots, X_{i-1})
  \label{eq-ChainRuleOfEntropy}
\end{equation}
where $H( X_{1} | \emptyset ) = H( X_{1} )$ for the empty set $\emptyset$.

Mutual information $I( X; Y)$ between probabilistic variables $X$ and $Y$ is defined as follows.
\begin{equation}
 I( X; Y) = H(X)+H(Y)-H(X, Y) = H(Y)-H(Y | X) = H(X) - H(X | Y) \ge 0
 \label{eq-MutualInformation}
\end{equation}
Conditional mutual information between probabilistic variable $X$ and $Y$ given $Z$ is defined as follows.
\begin{equation}
 I( X; Y | Z) = H( X | Z ) - H( X| Y, Z ) \ge 0
\label{eq-ConditionalMutualInformation}
\end{equation}
Nonnegativity of mutual information and conditional mutual information is given by Gibbs' inequality $\sum_{x}p(x)\log(p(x) - q(x)) \ge 0$ for any two probabilistic distributions $p(x) \ge 0$, $q(x) \ge 0$ and $\sum_{x}p(x) = \sum_{x}q(x)= 1$.
As a special case of Equation \ref{eq-ConditionalMutualInformation}, pairwise transfer entropy at time step $t$ is conditional mutual information between a series of variables $\bar{X}^{t}$
and $\bar{Y}^{t-1}$ given $\bar{X}^{t-1}$ as follows:
%\begin{equation}
 $T_{Y \rightarrow X}^{t} = I( \bar{X}^{t}; \bar{Y}^{t-1} | \bar{X}^{t-1})$
%\end{equation}
where $\bar{X}^{t} = \{ X^{1}, X^{2}, \hdots, X^{t}\}$, $Y = \bar{Y}^{t-1} = \{ X^{1}, X^{2}, \hdots, Y^{t-1}\}$.

{\it Total correlation} of the set of $N$ variables $\{X_{1}, X_{2}, \hdots, X_{N}\}$ 
is one of the multivariate extensions of mutual information proposed in \cite{Watanabe1960} (also known as {\it multivariate constraint} \cite{Garner1962} or {\it multiinformation} \cite{Studeny1999}).
\begin{equation}
 C( X_{1}, X_{2}, \hdots, X_{N} ) = \sum_{i}^{N}H(X_{i}) - H( X_{1}, X_{2}, \hdots, X_{N} )
  \label{eq-TotalCorrelation}
\end{equation}
Total correlation of two variables is obviously identical to mutual information.

Let $H( X_{\mathcal{S}} )$ be $H\left( X_{s_{1}}, X_{s_{2}}, \hdots, X_{s_{|\mathcal{S}|}} \right)$ for the set $\mathcal{S} = \left\{ s_{1}, s_{2}, \hdots, s_{|\mathcal{S}|} \right\}$.
Multivariate mutual information of $N$ variables $\{X_{1}, X_{2}, \hdots, X_{N}\}$ 
is another type of multivariate extension of mutual information
(also known as {\it interaction information} \cite{McGill1954}, {\it multiple mutual information} \cite{Han1978,Han1980}, {\it co-information} \cite{Bell2003}, and {\it synergy} \cite{Gawne1993}; see also \cite{Williams2010} for its review and generalization).
\begin{equation}
 I( X_{1}; X_{2}; \hdots; X_{N} ) = \sum_{k=1}^{N} (-1)^{k+1} \sum_{ \mathcal{S} \subset S( \mathcal{N}, k ) } H( X_{ \mathcal{S} } )
  \label{eq-MultivariateMutualInformation}
\end{equation}
where $\mathcal{N}$ is the set of indices $\{1, 2, \hdots, N\}$ and $S( \mathcal{N}, k ) = \{ \{ 1, 2, \hdots, k\}, \hdots, \{ N-k+1, N-k+2, \hdots, N \}\}$ is the combinatorial set including all the unique sets of size $k$ drawn from $\mathcal{N}$.
Then, multivariate conditional mutual information is naturally defined based upon multivariate mutual information.
\begin{equation}
 I( X_{1}; X_{2}; \hdots; X_{n} | X_{n+1} ) = I( X_{1}; X_{2}; \hdots; X_{n} ) - I( X_{1}; X_{2}; \hdots; X_{n+1} )
 \label{eq-MultivariateConditionalMutualInformation}
\end{equation}
Similarly with the chain rule of entropy (Equation \ref{eq-ChainRuleOfEntropy}), multivariate mutual information 
is the sum of conditional mutual information as follows \cite{CoverThomas1991}.
\begin{equation}
 I(Y; \{ X_{1}, X_{2}, \hdots, X_{n} \} ) = \sum_{i = 1}^{n} I( Y; X_{i}| \{ X_{1}, X_{2}, \hdots, X_{i-1} \})
  \label{eq-ChainRuleOfInformation}
\end{equation}

\section{ Information lattice }
In this section, we introduce the idea of {\it information lattice} which is defined for a set of random variables.
The information lattice gives an integrated view on statistical dependency among the set of random variables, and it is used 
in the proof of theoretical properties of multivariate transfer entropy.
Let us introduce a notation for the set of random variables $1, 2, \hdots, N$ each of which is defined on time index $1, 2, \hdots, T$ as follows.

Let $\bar{X}_{ \mathcal{N} }^{ \mathcal{T} }$ be a set of $N \times T$ probabilistic variables indexed with the index set for variables $\mathcal{N} = \{ 1, 2, \hdots, N \}$ and the index set for time $\mathcal{T} = \{1, 2, \hdots, T\}$.
Then let us denote
$\bar{X}_{ \mathcal{N} }^{ \mathcal{T} } = \{ \bar{X}_{1}^{ \mathcal{T} }, \bar{X}_{2}^{ \mathcal{T} }, \hdots, \bar{X}_{N}^{ \mathcal{T} } \} = \{ \bar{X}_{ \mathcal{N} }^{ 1 }, \bar{X}_{ \mathcal{N} }^{ 2 }, \hdots, \bar{X}_{ \mathcal{N} }^{ T } \}$ where 
$ \bar{X}_{i}^{ \mathcal{T} } = \{ X_{i}^{1}, X_{i}^{2}, \hdots, X_{i}^{T}\}$ denotes the time-cumulative subset of $X_{i}$ for time index
 $\mathcal{T}$, and 
 $ \bar{X}_{ \mathcal{N} }^{ t } = \{ X_{1}^{t}, X_{2}^{t}, \hdots, X_{ N }^{T} \}$ 
denotes a set for the variable set $\mathcal{N}$ given $t$.

%% Definition of information lattice
Then, the information lattice of $\bar{X}_{ \mathcal{N} }^{ \mathcal{T} }$ is defined as the set of multivariate conditional variables follows.
\begin{equation}
\tilde{I}(X_{1}^{ t_{1} }, X_{2}^{ t_{2} }, \hdots, X_{N}^{ t_{N} }) \equiv
   I\left( X^{t_{1}}_{1}; X^{t_{2}}_{2}; \hdots; X^{t_{N}}_{N} | 
     \bar{X}^{ \bar{t}_{1} \setminus t_{1} }_{1}, \bar{X}^{  \bar{t}_{2} \setminus t_{2}  }_{2}, \hdots, 
     \bar{X}^{ \bar{t}_{N} \setminus t_{N} }_{N} \right)
\label{eq-InformationLattice}
\end{equation}
where $\bar{t}_{i} = \{ 1, 2, \hdots, t_{i}\}$ for integer index $0 < t_{1}, t_{2}, \hdots, t_{N} \le T$, and in particular $X_{i}^{\emptyset} = \emptyset$.

The information lattice has the following theoretical property.

%%%
\begin{thm}[Partial expansion of information lattice]
\label{thm-PartialExpansionInformationLattice}
\begin{eqnarray}
    I\left( X^{\bar{t}_{1}}_{1}; X^{ \bar{t}_{2}}_{2}; \hdots; X^{ \bar{t}_{k}}_{k} | 
X^{ \bar{t}_{1} \setminus t_{1} }_{1}, 
\hdots 
X^{ \bar{t}_{k} \setminus t_{k} }_{k}, 
X^{ \bar{t}_{k+1} }_{k+1}, 
\hdots, X^{ \bar{t}_{N} }_{N} \right)
\\
=
\sum_{ m = 0 }^{ N - k }
(-1)^{m}\sum_{ s \subset S( \mathcal{N} \setminus \bar{k}, m )}
%\sum_{ t_{s} } \tilde{I}( \bar{t}_{k}, t_{s} )
 \sum_{i_{1}=1}^{t_{s_1}}
\hdots \sum_{i_{|s|}=1}^{t_{ s_{|s|} }}
\tilde{I}\left( X_{ 1 }^{ t_{1} }, \hdots X_{ k }^{ t_{k} }, X_{ s_1 }^{ i_{ s_1 } }, \hdots X_{ s_{ |s| } }^{ i_{ |s| } } 
\right) 
%\sum_{s_{1}=1}^{t_{1}}
%\hdots \sum_{s_{k}=1}^{t_{k}}
%\tilde{I}(s_{1}, \hdots s_{k}, t_{k+1}, \hdots, t_{N}) 
\label{eq-InformationLatticeProperty}
\end{eqnarray}
where $s \subset \{ s_{1}, s_{2}, \hdots, s_{ |s| }\}$ and $|s|$ is the size of the set $s$.
\end{thm}

%Due to the Theorem \ref{thm-PartialExpansionInformationLattice}, the following three properties 
\begin{cor}[Group-wise decomposition of joint entropy]
Equation \ref{eq-InformationLatticeProperty} in the special case with $k=1$ gives 
\begin{equation}
 H\left( X_{1}^{t_{1}} | X_{2}^{t_{2}}, \hdots, X_{N}^{t_{N}} \right) = 
\sum_{k=0}^{N-1} \sum_{ s \subset S( \mathcal{N} \setminus 1, k ) }
(-1)^{|s|}
\sum_{ i_{1} = 1 }^{ t_{s_1}} 
\hdots
\sum_{ i_{ |s| } = 1 }^{ t_{s_{|s|} }} 
\tilde{I}\left( X_{1}^{t_{1}}, X_{s_{1}}^{i_{s_{1}}}; \hdots; X_{s_{|s|}}^{i_{s_{|s|}}} \right)
\end{equation} 
Similarly, for $ 1 \le i \le N$
\begin{equation}
 H\left( X_{i}^{t_{i}} | X_{i+1}^{t_{i+1}}, \hdots, X_{N}^{t_{N}} \right) = 
\sum_{k=0}^{N-i} \sum_{ s \subset S( \mathcal{N} \setminus \bar{i}, k ) }
(-1)^{|s|}I\left( X_{i}^{t_{i}}; X_{s_{1}}^{t_{s_{1}}}; \hdots; X_{s_{|s|}}^{t_{s_{|s|}}} \right)
\end{equation} 
where $\bar{i} = \{ 1, 2, \hdots, i \}$.
Thus,
\begin{equation}
 H\left( X_{1}^{t_{1}}, X_{2}^{t_{2}}, \hdots, X_{N}^{t_{N}} \right) = 
\sum_{k=1}^{N} \sum_{ s \subset S( \mathcal{N}, k ) }
(-1)^{k+1}I\left( X_{s_{1}}^{t_{s_{1}}}; X_{s_{2}}^{t_{s_{2}}}; \hdots; X_{s_{|s|}}^{t_{s_{|s|}}} \right)
\end{equation} 
\end{cor}

\begin{cor}[Information lattice of a pair of variable groups]
\label{Cor-PairwiseInformationLattice}
Equation \ref{eq-InformationLatticeProperty} in the special case with $k=2$ gives 
\begin{equation}
 I\left( X_{i}^{t_{1}}; X_{2}^{t_{2}} | X_{3}^{t_{3}} \hdots, X_{N}^{t_{N}} \right) = 
\sum_{k=0}^{N-2} \sum_{ s \subset S( \mathcal{N} \setminus \{1, 2\}, k ) }
(-1)^{|s|}I\left( X_{1}^{t_{1}}; X_{2}^{t_{2}}; X_{s_{1}}^{t_{s_{1}}}; \hdots; X_{s_{|s|}}^{t_{s_{|s|}}} \right)
\end{equation} 
\end{cor}

 \begin{figure}[tb, clip]
   \begin{center}
    \includegraphics[width=1\linewidth, clip]{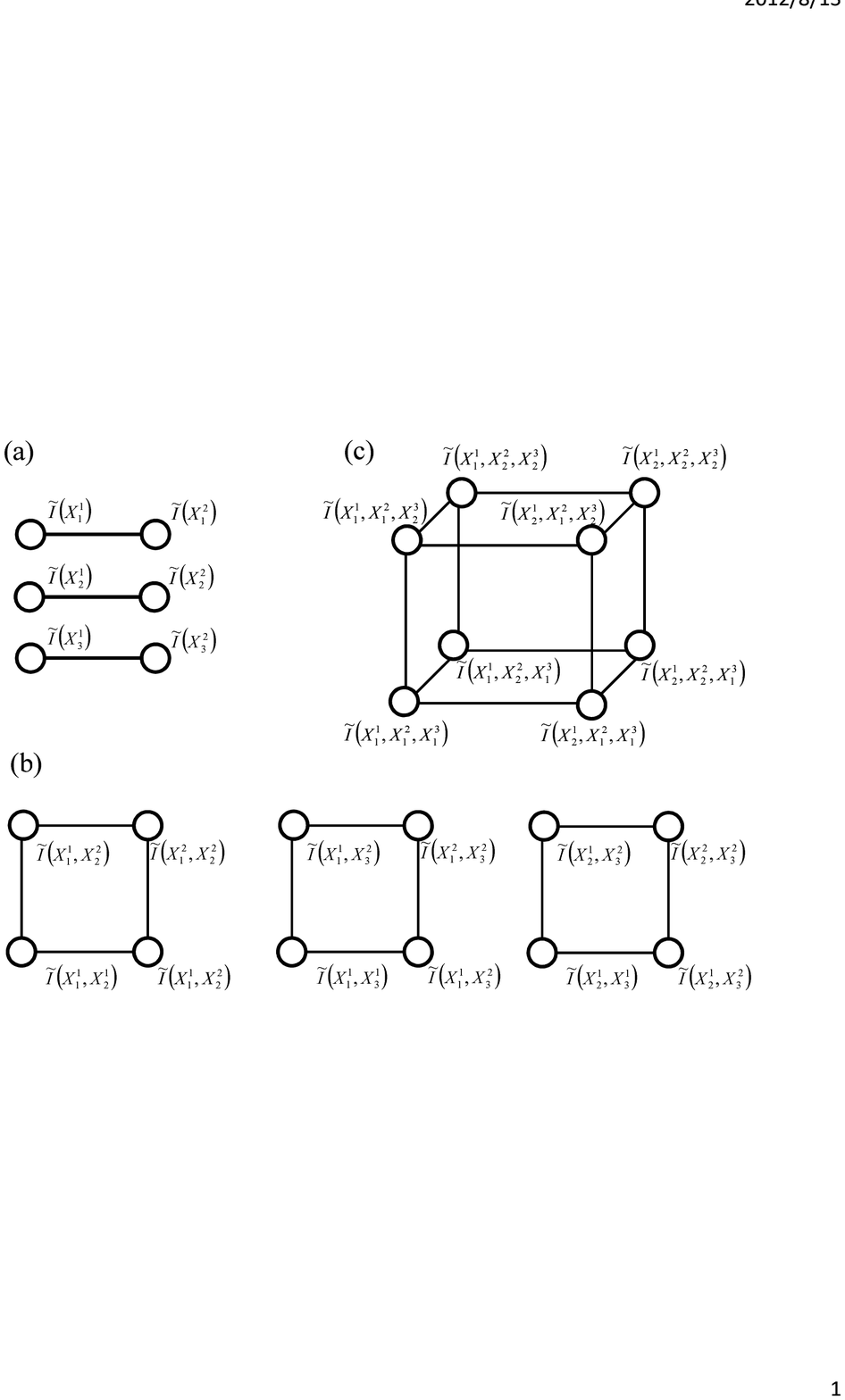}
    %\caption[]{Bidirectional information network between two variables $X$ and $Y$}  
    \caption{
Information lattice for three groups of random variables 
$X_{1} = \{ X_{1}^{1}, X_{1}^{2}\}$, $X_{2} = \{ X_{2}^{1}, X_{2}^{2}\}$, and $X_{3} = \{ X_{3}^{1}, X_{3}^{2}\}$.
(a) The first order, (b) the second order, and (c) the third order information lattice.
\label{fig-InformationLattice}}      
    \end{center}
  \end{figure}

\subsection{Proof of theoretical properties of information lattice}
%Three lemma was given prior to the proof of Theorem \ref{thm-PartialExpansionInformationLattice}.
Theorem \ref{thm-PartialExpansionInformationLattice} is naturally derived as one of the most general forms of combinations
among the three possible operations on the information lattices. Each of the operations is given Lemmas 1, 2 and 3, 
and Theorem \ref{thm-PartialExpansionInformationLattice} is derived as the result.
Three lemmas were given prior to the proof of Theorem \ref{thm-PartialExpansionInformationLattice}.
Lemma \ref{lem-IdentityMultivariateConditionalMutualInformation} shows the basic identity of multivariate conditional mutual information.
Lemma \ref{lem-ChainRuleOfConditionalMutualInformation} shows the chain rule of multivariate conditional mutual information
on basis of the basic identity Lemma \ref{lem-IdentityMultivariateConditionalMutualInformation}. This can be viewed as a generalization of the chain rule of entropy (Equation \ref{eq-ChainRuleOfEntropy})
and the chain rule of information (Equation \ref{eq-ChainRuleOfInformation}). Lemma \ref{thm-JointEntropyDecomposition} shows 
the relationship between joint entropy and its decomposition based on multivariate conditional mutual information
the using Lemma \ref{lem-ChainRuleOfConditionalMutualInformation}.
Theorem \ref{thm-PartialExpansionInformationLattice} showed all three properties, chain rule of multivariate conditional mutual information, sum of them, and the sum in form of the inclusion-excusion principle, in a general form. 
Thus, Theorem \ref{thm-PartialExpansionInformationLattice} is self-evident after the three lemmas.

%%%%%%%%%%%%%%%%%%%%%%%%%%%%%%%%%%%%%%%%%%%%%%%%%%%%%%%%%%%%%%%
%% Lemma 1: Identity of multivariate conditional mutual information
%%%%%%%%%%%%%%%%%%%%%%%%%%%%%%%%%%%%%%%%%%%%%%%%%%%%%%%%%%%%%%%
\begin{lem}[Identity of multivariate conditional mutual information]
\label{lem-IdentityMultivariateConditionalMutualInformation}
The following identity holds for multivariate mutual information.
\begin{eqnarray}
 %\left\{ %% THIS IS NUMERICALLY VALIDATED August 6th, 2012 by SHOHEI HIDAKA
\begin{array}{lll}
 I( X_{1}; X_{2}; \hdots; X_{n} | Y ) &=& I( X_{1}; X_{2}; \hdots; X_{n-1}; \{ X_{n}, Y \} ) - I( X_{1}; X_{2}; \hdots ; X_{n-1}; Y )
\\
 I( X_{1}; X_{2}; \hdots; X_{n} | Y ) &=& - I( X_{1}; X_{2}; \hdots; X_{n-1} | X_{n}, Y ) + I( X_{1}; X_{2}; \hdots ; X_{n-1} | Y )
 \label{eq-MultivariateMutualInformationConditioning}
\end{array}
%\right.
\end{eqnarray}

\begin{proof}
Let $G$ be $I( X_{1}; X_{2}; \hdots; X_{n-1}; \{ X_{n}, Y \} ) - I( X_{1}; X_{2}; \hdots; X_{n-1}; Y )$, then
\begin{eqnarray*}
 G &=& \sum_{k=1}^{N-1}(-1)^{k+1}\sum_{ \mathcal{S} \subset S( \mathcal{N} \setminus N, k ) } 
\left\{ ( H( \mathcal{S} ) - H( \mathcal{S}, X_{N}, Y) ) - ( H( \mathcal{S} ) - H( \mathcal{S}, Y) ) \right\}
\\
&=& \sum_{k=1}^{N-1}(-1)^{k+1}\sum_{ \mathcal{S} \subset S( \mathcal{N} \setminus N, k ) }\left\{  - H( \mathcal{S}, X_{N}, Y) +  H( \mathcal{S}, Y) \right\}
\\
&=& \sum_{k=1}^{N-1}(-1)^{k+1}\sum_{ \mathcal{S} \subset S( \mathcal{N} \setminus N, k ) }\left\{  - H( \mathcal{S}, X_{N}| Y) +  H( \mathcal{S}| Y) \right\}
\\
&=& \sum_{k=1}^{N-1}(-1)^{k+1}\sum_{ \mathcal{S} \subset S( \mathcal{N} \setminus N, k ) }  H( \mathcal{S} | Y )
= I( X_{1}; X_{2}; \hdots; X_{n} | Y )
%I( X_{1}; X_{2}; \hdots; \{ X_{n}, Y \} ) - I( X_{1}; X_{2}; \hdots; Y )
\end{eqnarray*}
\end{proof}
\end{lem}

%%%%%%%%%%%%%%%%%%%%%%%%%%%%%%%%%%%%%%%%%%%%%%%%%%%%%%%%%%%%%%%
%% Lemma 2: Chain rule of conditional mutual information
%%%%%%%%%%%%%%%%%%%%%%%%%%%%%%%%%%%%%%%%%%%%%%%%%%%%%%%%%%%%%%%
\begin{lem}[Chain rule of multivariate conditional mutual information]
\label{lem-ChainRuleOfConditionalMutualInformation}
Conditional mutual information is decomposed with the sum of conditional mutual information as follows.
\begin{eqnarray}
%I\left( \bar{X}_{\mathcal{N} \setminus N }^{T}; \bar{X}_{N}^{ \mathcal{T}_{N} }
%| \bar{X}^{ \mathcal{T} \setminus  T  }_{ \mathcal{N} \setminus N } \right)
I\left( \bar{X}_{1}^{ \mathcal{T} }; \bar{X}_{2}^{ \mathcal{T} }; \hdots ; \bar{X}_{N}^{ \mathcal{T}_{N} }
| \bar{X}^{ \mathcal{T} \setminus  T  }_{ \mathcal{N} \setminus N } \right)
= \sum_{ t_{N} = 1}^{ T_{N} } 
%I\left( \bar{X}_{ \mathcal{N} \setminus N }^{T}; \bar{X}_{ N }^{ t_{N} } | \bar{X}^{ \mathcal{T} \setminus T }_{ \mathcal{N} \setminus N}, \bar{ X }_{N}^{ \bar{t}_{N} \setminus t_{N} } \right)
I\left( 
\bar{X}_{1}^{ \mathcal{T} }; \bar{X}_{2}^{ \mathcal{T} }; \hdots ; \bar{X}_{N-1}^{ \mathcal{T} }; \bar{X}_{ N }^{ t_{N} } | \bar{X}^{ \mathcal{T} \setminus T }_{ \mathcal{N} \setminus N}, \bar{ X }_{N}^{ \bar{t}_{N} \setminus t_{N} } \right)
\label{eq-LemmaMultivariateConditionalMutualInformation1}
\end{eqnarray}
where $\bar{t_{N}} = \{ 1, 2, \hdots, t_{N}\}$.
We also define $ I( \emptyset; X | Y) = I( X ; \emptyset | Y) \equiv H(X | Y)$ for convenience.
Then, $ \bar{X}_{ \mathcal{N} \setminus N }^{T} = \emptyset$ ascribes Equation \ref{eq-LemmaMultivariateConditionalMutualInformation1}
to the chain rule of entropy (Equation \ref{eq-ChainRuleOfEntropy})
\begin{eqnarray}
H\left( \bar{X}_{N}^{ \mathcal{T}_{N} } \right)
= \sum_{ t_{N} = 1}^{ T_{N} } H\left( X_{ N }^{ t_{N} } | \bar{X}_{N}^{ t_{N}-1} \right)
\end{eqnarray}

Equivalent to Equation (\ref{eq-LemmaMultivariateConditionalMutualInformation1}) above,
%\begin{eqnarray} 
%G = I\left( \bar{X}_{ \mathcal{N} \setminus \{ N \} }^{ T } | X^{ \mathcal{T} \setminus \{ T \} }_{ \mathcal{N} \setminus \{ N \} }, X_{N}^{ %\mathcal{T}_{N} } \right)
%\label{eq-LemmaMultivariateConditionalMutualInformation2}
%\end{eqnarray}
where
%$G = I\left( x^{T}_{1}; x^{T}_{2}; \hdots; x^{T}_{N-1} | X^{ \mathcal{T}\setminus T }_{ \mathcal{N} \setminus N } \right)
%- \sum_{ t_{N} = 1}^{ T_{N} } I\left( x_{1}^{T}; x_{2}^{T}; \hdots; x_{N-1}^{T}; x_{ N }^{ t_{N} } | \bar{X}^{ \{ \mathcal{T} \setminus T \} }_{ \mathcal{N} \setminus N}, \bar{X}_{N}^{\{ 1, 2, \hdots, t_{N}-1 \}} \right)$. 
\begin{equation}
\label{eq-LemmaMultivariateConditionalMutualInformation2}
I\left( \bar{X}_{ \mathcal{N} \setminus  N  }^{ T } | \bar{X}^{ \mathcal{T} \setminus T  }_{ \mathcal{N} \setminus N }, \bar{X}_{N}^{ \mathcal{T}_{N} } \right)
=
%\bar{I}\left( \bar{X}^{T}_{ \mathcal{N} } | \bar{X}^{\mathcal{T} \setminus T}_{ \mathcal{N} } \right)
%- \sum_{ t_{N} = 1}^{ T_{N} } I\left( X_{1}^{T}; X_{2}^{T}; \hdots; X_{N-1}^{T}; X_{ N }^{ t_{N} } | \bar{X}^{ \mathcal{T} \setminus T }_{ %\mathcal{N} \setminus N}, \bar{X}_{N}^{ \bar{t}_{N} \setminus t_{N} } \right)
\tilde{I}(t_{1}, t_{2}, \hdots, t_{N-1})
-
\sum_{t_{N} = 1}^{T_{N}}\tilde{I}(t_{1}, t_{2}, \hdots, t_{N})
\end{equation}
where
partial expansion of multivariate conditional mutual information %$\tilde{I}(t_{1}, t_{2}, \hdots, t_{N})$ 
is defined as follows.
\begin{equation}
\tilde{I}(t_{1}, t_{2}, \hdots, t_{N}) = 
   I\left( X^{t_{1}}_{1}; X^{t_{2}}_{2}; \hdots; X^{t_{N}}_{N} | 
     \bar{X}^{ \bar{t}_{1} \setminus t_{1} }_{1}, \bar{X}^{  \bar{t}_{2} \setminus t_{2}  }_{2}, \hdots, 
     \bar{X}^{ \bar{t}_{N} \setminus t_{N} }_{N} \right)
\label{eq-PartialDecompositionMultivariateConditionalMutualInformation}
\end{equation}
%
%
%where $\bar{I}\left( \bar{X}^{T}_{ \mathcal{N} } | \bar{X}^{\mathcal{T} \setminus T}_{ \mathcal{N} } \right) 
%= I\left( {X}^{T}_{1}; {X}^{T}_{2}; \hdots; {X}^{T}_{N} | X^{ \mathcal{T}\setminus T }_{ \mathcal{N} } \right)$.
%{ \bf TODO }
%Also let $\bar{I}(X^{ \mathcal{T} }_\mathcal{N})$ be sum of mutual condition over time 
%$\sum_{ t = 1}^{ T } I \left( X^{t}_\mathcal{N} | X^{ \{ \mathcal{T} - 1 \} }_\mathcal{N} \right) 
%= \sum_{ t = 1}^{T} I\left( X_{t}^{1}; X_{t}^{2}; \hdots; X_{t}^{N} | X^{ \{ 1, 2, \hdots, t - 1 \} }_\mathcal{N} \right) $.
%In a special case with a single variable, it ascribes to cumulative sum of entropy rate $\bar{I}\left( \bar{X}^{ \mathcal{T} }_i | \bar{X}^{ %\mathcal{T} \setminus T }_i \right) = \sum_{t=1}^{
%T } H \left( x^{t}_{i}| \bar{X}^{ \{1, 2, \hdots, t-1\} }_{i} \right)$.
%
Applying iteratively the chain rule of multivariate conditional mutual information (Equation \ref{eq-LemmaMultivariateConditionalMutualInformation2}), 
 \begin{equation}
  I\left( \bar{X}_{1}^{\mathcal{T}}; \bar{X}_{2}^{\mathcal{T}}; \hdots, \bar{X}_{N}^{\mathcal{T}} \right) 
   = \sum_{ t_{1} = 1 }^{T} 
   \sum_{ t_{2} = 1 }^{T} 
   \hdots
   \sum_{ t_{N} = 1 }^{T} 
   \tilde{I}(t_{1}, t_{2}, \hdots, t_{N}) 
\label{eq-PartialDecompositionMultivariateMutualInformation}
%  \label{lem-MultivariateMutualInformationIdentity}
 \end{equation}

%%%%%%%%%%%%%%%%%%%%%%%%%%%%%%%%%%%%%%%%%%%%%%%%%%%%%%%%%%%%%%%
%% Proof of Lemma 2: Chain rule of conditional mutual information
%%%%%%%%%%%%%%%%%%%%%%%%%%%%%%%%%%%%%%%%%%%%%%%%%%%%%%%%%%%%%%%

\begin{proof}
Let $G$ be the right hand side of Equation \ref{eq-LemmaMultivariateConditionalMutualInformation2}, and 
applying Equation (\ref{eq-MultivariateMutualInformationConditioning}) 
to $G$ repeatedly,
\begin{eqnarray*}
% \bar{I}( X_{ \mathcal{T} }^{ \mathcal{N} \setminus N } ) - \sum_{ t = 1 }^{T} \bar{I}( X^{ \mathcal{T} }_{ \mathcal{N} \setminus N } ) 
G &=& 
I\left( \bar{X}_{\mathcal{N} \setminus \{ N \} }^{T} 
%X^{t}_{1}; X^{t}_{2}; \hdots; X^{t}_{N-1} 
| X^{ \mathcal{T} \setminus \{ T \} }_{ \mathcal{N} \setminus \{ N \} } \right)
- \sum_{ t_{N} = 1}^{ T_{N} } I\left( \bar{X}_{ \mathcal{N} - 1 }^{T}; x_{ N }^{ t_{N} } | X^{ \{ \mathcal{T} \setminus T \} }_{ \mathcal{N} \setminus N}, X_{N}^{ t_{N}-1} \right)
%\sum_{ t = 1}^{T} I \left( X^{t}_\mathcal{N} | X^{ \{ 1 , \hdots, t-1 \} }_\mathcal{N} \right) 
\\
&=& 
I\left( \bar{X}_{\mathcal{N} \setminus \{ N \} }^{T} 
%X^{t}_{1}; X^{t}_{2}; \hdots; X^{t}_{N-1} 
| X^{ \mathcal{T} \setminus \{ T \} }_{ \mathcal{N} \setminus \{ N \} }, x_{N}^{1} \right)
- \sum_{ t_{N} = 2}^{ T_{N} } I\left( \bar{X}_{ \mathcal{N} - 1 }^{T}; x_{ N }^{ t_{N} } | X^{ \{ \mathcal{T} \setminus T \} }_{ \mathcal{N} \setminus N}, X_{N}^{ t_{N}-1} \right)
%\sum_{ t = 1}^{T} I \left( X^{t}_\mathcal{N} | X^{ \{ 1 , \hdots, t-1 \} }_\mathcal{N} \right) 
\\
&=& 
I\left( \bar{X}_{\mathcal{N} \setminus \{ N \} }^{T} 
%X^{t}_{1}; X^{t}_{2}; \hdots; X^{t}_{N-1} 
| X^{ \mathcal{T} \setminus \{ T \} }_{ \mathcal{N} \setminus \{ N \} }, x_{N}^{\{ 1, 2, \hdots t'_{N}\}} \right)
- \sum_{ t_{N} = t'_{N} + 1}^{ T_{N} } I\left( \bar{X}_{ \mathcal{N} - 1 }^{T}; x_{ N }^{ t_{N} } | X^{ \{ \mathcal{T} \setminus T \} }_{ \mathcal{N} \setminus N}, X_{N}^{ t_{N}-1} \right)
%\sum_{ t = 1}^{T} I \left( X^{t}_\mathcal{N} | X^{ \{ 1 , \hdots, t-1 \} }_\mathcal{N} \right) 
\\
&=&  
I\left( \bar{X}_{ \mathcal{N} \setminus \{ N \} }^{ T } | X^{ \mathcal{T} \setminus \{ T \} }_{ \mathcal{N} \setminus \{ N \} }, X_{N}^{ \mathcal{T}_{N} } \right)
\end{eqnarray*}
Thus, Equation \ref{eq-LemmaMultivariateConditionalMutualInformation1} is proven.
\end{proof}

\end{lem}
%%%%%%%%%%%%%%%%%%%%%%%%%%%%%%%%%%%%%%%%%%%%%%%%%%%%%%%%%%%%%%%
%% END of Lemma 2: Chain rule of conditional mutual information
%%%%%%%%%%%%%%%%%%%%%%%%%%%%%%%%%%%%%%%%%%%%%%%%%%%%%%%%%%%%%%%

%\subsection{ Decomposition of joint entropy with information }
%%%%%%%%%%%%%%%%%%%%%%%%%%%%%%%%%%%%%%%%%%%%%%%%%%%%%%%%%%%%%%%
%% Lemma 3: Decomposition of joint entropy with information
%%%%%%%%%%%%%%%%%%%%%%%%%%%%%%%%%%%%%%%%%%%%%%%%%%%%%%%%%%%%%%%
\begin{lem}[Decomposition of joint entropy with information]
 For $N$ series of probabilistic variables $ X^{ \mathcal{T}  }_{ \mathcal{N }} $, 
 joint entropy of the $N$ variables $ H( X^{ \mathcal{T} }_{\mathcal{N}} )$ can be decomposed as follows.
 \label{thm-JointEntropyDecomposition}
 \begin{eqnarray}
  H\left( X_{ \mathcal{T} }^{\mathcal{N}} \right) 
   = 
   \sum_{k = 1}^{ N } (-1)^{ k + 1 } \sum_{ \mathcal{S} \subset S( \mathcal{N}, k ) }
   I\left( \bar{X}_{ \mathcal{S} }^{ \mathcal{T} } \right) 
%   = 
%   \sum_{k = 1}^{ N } (-1)^{ k + 1 } \sum_{ s \subset S( \mathcal{N}, k ) }
%   \bar{I}( s, \mathcal{T}  ) 
   \label{eq-JointEntropyDecomposition}
 \end{eqnarray}
 where $I$ is partial decomposition of multivariate mutual information (Equation \ref{eq-PartialDecompositionMultivariateConditionalMutualInformation}) as follows.
 \begin{equation}
  I\left( \bar{X}_{ \mathcal{S} }^{ \mathcal{T} } \right) 
 \equiv
 I( \bar{X}_{ 1 }^{ \mathcal{T} }; \bar{X}_{ 2 }^{ \mathcal{T} }; \hdots; \bar{X}_{ N }^{ \mathcal{T} } ) 
      %\bar{I}( \mathcal{N}, \mathcal{T} ) 
   = \sum_{ t_{1} = 1 }^{T} 
   \sum_{ t_{2} = 1 }^{T} 
   \hdots
   \sum_{ t_{N} = 1 }^{T} 
   \tilde{I}(t_{1}, t_{2}, \hdots, t_{N}) 
   %I\left( x^{t_{1}}_{1}; x^{t_{2}}_{2}; \hdots; x^{t_{N}}_{N} | 
   %  \bar{X}^{ \bar{t}_{1} \setminus t_{1} }_{1}, \bar{X}^{  \bar{t}_{2} \setminus t_{2}  }_{2}, \hdots, 
   %  \bar{X}^{ \bar{t}_{N} \setminus t_{N} }_{N} \right)
 \end{equation}

%%%%%%%%%%%%%%%%%%%%%%%%%%%%%%%%%%%%%%%%%%%%%%%%%%%%%%%%%%%%%%%
%% Proof of Lemma 3: Decomposition of joint entropy with information
%%%%%%%%%%%%%%%%%%%%%%%%%%%%%%%%%%%%%%%%%%%%%%%%%%%%%%%%%%%%%%%
\begin{proof} %[Decomposition of joint entropy with information]
Applying iteratively the chain rule of conditional mutual information 
(Equation \ref{eq-LemmaMultivariateConditionalMutualInformation2}), we obtain Lemma 
 \ref{thm-JointEntropyDecomposition} as follows. %and Equation (\ref{eq-FreeEntropyNVariables}),
\begin{eqnarray}
 H( \bar{X}_{ \mathcal{T} }^{\mathcal{N}} ) 
%\nonumber
&=&
\sum_{ i = 1 }^{ N } H( \bar{X}_{i}^{\mathcal{T}} | \bar{X}_{ \bar{i} \setminus i }^{\mathcal{T}} )
\label{eq-DecompositionJointEntropyMiddle1}
\\
&=&
\sum_{ i = 1 }^{ N } \left\{ 
H( \bar{X}_{i}^{\mathcal{T}} | X_{ \bar{i} \setminus i }^{\mathcal{T}} )
\pm
H( \bar{X}_{i}^{\mathcal{T}} | \bar{X}_{ \bar{i} \setminus \{ i, i-1\} }^{\mathcal{T}} )
\pm
H( \bar{X}_{i}^{\mathcal{T}} | \bar{X}_{ \bar{i} \setminus \{ i, i-1, i-2\} }^{\mathcal{T}} )
\hdots
\pm
H( \bar{X}_{i}^{\mathcal{T}}  )
\right\}
\nonumber
\\
&=& 
\label{eq-DecompositionJointEntropyMiddle2}
\sum_{ i = 1 }^{ N }
\left\{ 
 H \left( \bar{X}_{ i }^{ \mathcal{T} }  \right)
 - \sum_{ j }^{ i - 1 }
 I\left( \bar{X}_{ i }^{ \mathcal{T} }; \bar{X}_{ j }^{ \mathcal{T} } | \bar{X}_{ \bar{ j } \setminus j }^{ \mathcal{T} } \right) 
\right\}
%\label{eq-JointEntropyDecompositionMiddle}
\\
&=& 
\sum_{ i = 1 }^{ N }
\left\{ 
 H \left( \bar{X}_{ i }^{ \mathcal{T} }  \right)
 - \sum_{ j }^{ i - 1 }
\left\{
 I\left( \bar{X}_{ i }^{ \mathcal{T} }; \bar{X}_{ j }^{ \mathcal{T} } | \bar{X}_{ \bar{ j } \setminus j }^{ \mathcal{T} } \right) 
\pm
 I\left( \bar{X}_{ i }^{ \mathcal{T} }; \bar{X}_{ j }^{ \mathcal{T} } | \bar{X}_{ \bar{ j } \setminus \{ j, j-1 \} }^{ \mathcal{T} } \right) 
\hdots
\pm I\left( \bar{X}_{ i }^{ \mathcal{T} }; \bar{X}_{ j }^{ \mathcal{T} } \right) 
\right\}
\right\}
\nonumber
\\
&=& 
\sum_{ i = 1 }^{ N }
\left\{ 
 H \left( \bar{X}_{ i }^{ \mathcal{T} }  \right)
 - \sum_{ j }^{ i - 1 }
\left\{
 I\left( \bar{X}_{ i }^{ \mathcal{T} }; \bar{X}_{ j }^{ \mathcal{T} } \right) 
-
\sum_{ k = 1}^{j-1}
 I\left( \bar{X}_{ i }^{ \mathcal{T} }; \bar{X}_{ j }^{ \mathcal{T} }; \bar{X}_{ k }^{ \mathcal{T} } | \bar{X}_{ \bar{ k } \setminus k }^{ \mathcal{T} } \right) 
\right\}
\right\}
\nonumber
\\
& & \vdots \nonumber
\\
&=& \sum_{ k = 1}^{N} (-1)^{k+1}\sum_{ \mathcal{S} \subset S(\mathcal{N}, k)} I( X_{ \mathcal{S} }^\mathcal{T} )  
%\\
%&=& \sum_{ k = 1}^{N} (-1)^{k+1}\sum_{ s \subset S(\mathcal{N}, k)} \bar{I}( s, \mathcal{T} )  
\nonumber
\end{eqnarray}
where $\bar{i} = \{ 1, 2, \hdots, i \}$, $\bar{j} = \{ 1, 2, \hdots, j \}$, and $\bar{k} = \{ 1, 2, \hdots, k \}$.
%and we applied identity (\ref{eq-MultivariateMutualInformationIdentity}) to the last line.
\end{proof}
%%%%%%%%%%%%%%%%%%%%%%%%%%%%%%%%%%%%%%%%%%%%%%%%%%%%%%%%%%%%%%%%%%%%%%
\end{lem}
%%%%%%%%%%%%%%%%%%%%%%%%%%%%%%%%%%%%%%%%%%%%%%%%%%%%%%%%%%%%%%%
%% END of Lemma 3: Decomposition of joint entropy with information
%%%%%%%%%%%%%%%%%%%%%%%%%%%%%%%%%%%%%%%%%%%%%%%%%%%%%%%%%%%%%%%
\begin{cor}
\label{cor-TotalCorrelation}
With Lemma \ref{thm-JointEntropyDecomposition}, the total correlation 
of the $N$ variables $C( \bar{X}^{\mathcal{T}}_{\mathcal{N}} )$ can be decomposed as follows.
\begin{equation}
 C\left( X_{T}^{\mathcal{N}} \right) 
 = \sum_{i=1}^{N}H\left( \bar{X}^{ \mathcal{T} }_{i} \right) 
%-
%\sum_{ k = 1}^{N} (-1)^{ k + 1 }\sum_{ \mathcal{S} \subset S(\mathcal{N}, k)} \bar{I}( \mathcal{S}, \mathcal{T} ) 
- \sum_{ k = 1}^{N} (-1)^{k+1}\sum_{ \mathcal{S} \subset S(\mathcal{N}, k)} I( X_{ \mathcal{S} }^\mathcal{T} )  
 \label{eq-TotalCorrelationNVariables2}
\end{equation}
Therefore, Lemma \ref{thm-JointEntropyDecomposition} explicitly states the mathematical relationship between joint entropy
and multivariate mutual information which is dual to Equation \ref{eq-MultivariateMutualInformation}.
\end{cor}
%%%%%%%%%%%%%%%%%%%%%%%%%%%%%%%%%%%%%%%%%%%%%%%%%%%%%%%%%%%%%%%

\begin{cor}
Applying the same reasoning in Lemma \ref{thm-JointEntropyDecomposition}, decomposition of residual entropy (Equation \ref{eq-ResidualEntropyN}) is given as follows.
\begin{eqnarray}
R_{ i, j }^{T}
& = & 
\sum_{t=1}^{T} I\left( X_{i}^{t}; X_{j}^{t} |  \bar{X}_{ \{ i, j \} }^{ \bar{t} \setminus t }, \bar{X}_{ \mathcal{N} \setminus \{ i, j \} }^{t} \right)
%\\
%& = & 
%\sum_{k=0}^{N-2}(-1)^{k}
%\sum_{ \mathcal{S} \subset S(\mathcal{N} \setminus \{ i, j \}, k )}
%\sum_{t = 1}^{T} %\Big\{ %\sum_{k = 2}^{N}
%I\left( 
%X_{i}^{t}; X_{j}^{t}; \bar{X}_{ \mathcal{S} }^{t}| X_{ \{ i, j \} }^{ \bar{t} \setminus t }, X_{ \mathcal{S} }^{ \bar{t} \setminus t }  
%\right)
\\
& = & 
( -1 )^{N-2}
\sum_{k=0}^{N-2}(-1)^{k}
\sum_{ \mathcal{S} \subset S(\mathcal{N} \setminus \{ i, j \}, k )}
\sum_{t = 1}^{T} %\Big\{ %\sum_{k = 2}^{N}
I\left( 
X_{i}^{t}; X_{j}^{t}; X_{s_{1}}^{t}; \hdots; X_{s_{|\mathcal{S}|}}^{t} | X_{ \{ i, j \} }^{ \bar{t} \setminus t }, X_{ \mathcal{S} }^{ \bar{t} \setminus t }  
\right)
\\
& = & 
\sum_{k=0}^{N-2}(-1)^{N - k}
\sum_{ \mathcal{S} \subset S(\mathcal{N} \setminus \{ i, j \}, k )}
\sum_{t = 1}^{T} %\Big\{ %\sum_{k = 2}^{N}
I\left( 
X_{i}^{t}; X_{j}^{t}; X_{s_{1}}^{t}; \hdots; X_{s_{|\mathcal{S}|}}^{t} | X_{ \{ i, j \} }^{ \bar{t} \setminus t }, X_{ \mathcal{S} }^{ \bar{t} \setminus t }  
\right)
\end{eqnarray}
where $\mathcal{S} \supset \{ s_{1}, s_{2}, \hdots, s_{ |\mathcal{S}| }\}$.

%%%%%%%%%%%%%%%%%%%%%%%%%%%%%%%%%%%%%%%%%%%%%%%%%%%%%%%%%%%%%%%%%%%%%%%%%%%%%%%%%%%
%%% Corollary on the residual entropy
%%%%%%%%%%%%%%%%%%%%%%%%%%%%%%%%%%%%%%%%%%%%%%%%%%%%%%%%%%%%%%%%%%%%%%%%%%%%%%%%%%%
\end{cor}

\section{Multivariate bidirectional information network}
%Let $\bar{X}_{ \mathcal{N} }^{ \mathcal{T} }$ be a set of $N \times T$ probabilistic variables indexed with the index set for variables $\mathcal{N} = \{ 1, 2, \hdots, N \}$ and the index set for time $\mathcal{T} = \{1, 2, \hdots, T\}$.
%Then let us denote
%$\bar{X}_{ \mathcal{N} }^{ \mathcal{T} } = \{ \bar{X}_{1}^{ \mathcal{T} }, \bar{X}_{2}^{ \mathcal{T} }, \hdots, \bar{X}_{N}^{ \mathcal{T} } \} = \{ \bar{X}_{ \mathcal{N} }^{ 1 }, \bar{X}_{ \mathcal{N} }^{ 2 }, \hdots, \bar{X}_{ \mathcal{N} }^{ T } \}$ where 
%$ \bar{X}_{i}^{ \mathcal{T} } = \{ X_{i}^{1}, X_{i}^{2}, \hdots, X_{i}^{T}\}$ denotes time-cumulative subset of $X_{i}$ for time index
% $\mathcal{T}$, and 
% $ \bar{X}_{ \mathcal{N} }^{ t } = \{ X_{1}^{t}, X_{2}^{t}, \hdots, X_{ N }^{T} \}$ 
%denotes a set for the variable set $\mathcal{N}$ given $t$.

Given the spatiotemporal set of probabilistic variables $\bar{X}_{ \mathcal{N} }^{ \mathcal{T} }$, entropy rate , free entropy, conditional transfer entropy and redundant entropy are defined as follows.
Entropy rate of variable $i$ at time $T$ is
\begin{eqnarray}
H_{i}^{T} \equiv %H \left( x_{N}^{ T } | \bar{X}_{ N }^{ \mathcal{T} \setminus T }  \right)
%H \left( x_{ i }^{ T } | \bar{X}_{ i }^{ \mathcal{T} \setminus T }  \right)
H \left( \bar{ X }_{ i }^{ \mathcal{T} }  \right)
= \sum_{t = 1}^{ T } H \left( X_{i}^{t} | \bar{ X }_{ i }^{ \bar{t} \setminus t }  \right)
\label{eq-EntropyRateN}
\end{eqnarray}
where $\bar{t} = \{1, 2, \hdots, t\}$ is the cumulative set of time indices 
and $\bar{t} \setminus t = \{1, 2, \hdots, t-1\}$ means set subtraction of index $t$ from $\bar{t}$.
The rightmost equation in entropy rate (\ref{eq-EntropyRateN}) is derived using the chain rule of entropy (Equation \ref{eq-ChainRuleOfEntropy}).

Free entropy of variable $i$ at time $t$ in $N$-variable system, which is uncertainty of $x_{i}$ given all the past states of $N$ variables $X^{ \mathcal{T} \setminus T }_{\mathcal{N}}$, is defined as follows.
\begin{eqnarray}
F_{i}^{T} \equiv 
\sum_{t=1}^{T}
H \left( X^{t}_{i}| X^{ \bar{t} \setminus t }_{\mathcal{N}} \right)
= 
\sum_{t=1}^{T}
\left\{
H \left( \bar{X}_{i}^{ \bar{t} }, \bar{X}_{ \mathcal{N} \setminus i}^{ \bar{t} \setminus t }  \right) 
- H \left(  \bar{X}^{ \bar{t} \setminus t }_{\mathcal{N}} \right)
\right\}
\label{eq-FreeEntropyN}
\end{eqnarray}
%%
%% Definition of Conditional transfer entropy[
%%
Then, conditional transfer entropy from variable $j$ to $i$ given the set of the other variables $\{ \mathcal{N} \setminus \{ i, j \} \}$ 
in the $N$-variable system $X^{ \mathcal{T} \setminus T }_{\mathcal{N}}$ is 
defined as follows.
\begin{eqnarray}
 T_{ j \rightarrow i | \mathcal{N} \setminus \{ i, j \} }^{T} 
\equiv
\sum_{t}^{T} I\left( X_{i}^{t}; \bar{X}_{j}^{ \bar{t} \setminus t } | \bar{X}_{ \mathcal{N} \setminus j }^{ \bar{t} \setminus \{t-1, t\} } \right)
=
  \sum_{t = 1}^{T} \sum_{ s = 1 }^{t-1}  I \left( X^{t}_{i}; X^{s}_{j} |\bar{X}^{ \bar{t} \setminus t }_{i}, \bar{X}^{ \bar{s} \setminus s }_{j}, X^{ \bar{ s } \setminus s }_{\mathcal{N} \setminus \{ i, j \}} \right)
  \label{eq-MultivariateTransferEntropy}
\end{eqnarray}
The far right hand side of Equation \ref{eq-MultivariateTransferEntropy} is derived with Lemma \ref{lem-ChainRuleOfConditionalMutualInformation}.

%%
%% Definition of Residual entropy
%%
%
%% Global residual
Residual information from variable $i$ in a $N$-variable system $X^{ \mathcal{T} \setminus T }_{\mathcal{N}}$ is defined as follows.
\begin{eqnarray}
R^{T}
 \equiv
\sum_{t=1}^{T} I\left( X_{1}^{t}; X_{2}^{t}; \hdots; X_{N}^{t} |  \bar{X}_{ \mathcal{N} }^{\bar{t} \setminus t} \right)
\label{eq-ResidualEntropyGlobal}
\end{eqnarray}
%% 
%% Single residual
%%
\begin{eqnarray}
R_{i}^{T}
 \equiv
\sum_{t=1}^{T} H\left( X_{i}^{t} |  \bar{X}_{ \mathcal{N} \setminus i }^{\bar{t} \setminus t} \right)
- \sum_{ j \subset S( \mathcal{N} \setminus i, 1)}
T_{ j \rightarrow i | \mathcal{N} \setminus \{ i, j \} }^{T}
\label{eq-ResidualEntropyGlobal}
\end{eqnarray}
%%
%% Pairwise residual
%% 
Residual information from variable $i$ in a $N$-variable system $X^{ \mathcal{T} \setminus T }_{\mathcal{N}}$ is defined as follows.
\begin{eqnarray}
R_{ i, j }^{T}
 \equiv
\sum_{t=1}^{T} I\left( X_{i}^{t}; X_{j}^{t} |  \bar{X}_{ \mathcal{N} }^{\bar{t} \setminus t} \right)
= 
\sum_{k=0}^{N-2}(-1)^{k}
\sum_{ \mathcal{S} \subset S(\mathcal{N} \setminus \{ i, j \}, k )}
R_{ i, j | \mathcal{S} }^{T} 
\label{eq-ResidualEntropyN}
\end{eqnarray}
where, for $\mathcal{S} \supset \{ s_{1}, s_{2}, \hdots, s_{ | \mathcal{S} | }\}$,
\begin{eqnarray}
R^{T}_{ i, j | \mathcal{S} } 
=
\sum_{t = 1}^{T}
I\left( 
%X_{i}^{t}; X_{j}^{t}; X_{s_{1}}^{t}; \hdots; X_{s_{|\mathcal{S}|}}^{t} | X_{ \{ i, j \} }^{ \bar{t} \setminus t }, X_{ \mathcal{S} }^{ \bar{t} \setminus t }  
X_{i}^{t}; X_{j}^{t}; \bar{X}_{s_{1}}^{\bar{t}}; \hdots; \bar{X}_{s_{|\mathcal{S}|}}^{ \bar{t} } | X_{ \{ i, j \} }^{ \bar{t} \setminus t }
\right)
\label{eq-ConditionalResidualEntropyN}
\end{eqnarray}.

Obviously, each of the entropies and informations are non-negative: $H_{i} \ge 0$, $F_{i} \ge 0$, $T_{j \rightarrow i | \mathcal{N} \setminus \{i, j\}} \ge 0$, and $R_{ i, j } \ge 0$ for arbitrary $ i \subset \mathcal{N}$ and $ j \subset \mathcal{ N } \setminus i$ 
because they are the sum of either conditional entropy (Equation \ref{eq-ConditionalEntropy}) or conditional mutual information (Equation \ref{eq-ConditionalMutualInformation}).
Although it is not obvious, each of residual entropy $R_{i}^{T}$ alone or the sum of residual entropy $(R_{i}^{T} + R^{T})$ is also non-negative which is shown later.

\subsection{Theoretical properties of multivariate bidirectional information network}
%%%
%%% In-middle outline of proof
%%% 
%\subsection{Proof of Theorem \ref{thm-TotalCorrelationDecompositionNVariables} and \ref{thm-LocalInformationFlow}}
Given multivariate information network defined with Equation 
\ref{eq-EntropyRateN}, \ref{eq-FreeEntropyN}, \ref{eq-MultivariateTransferEntropy}, and \ref{eq-ConditionalResidualEntropyN}, the two important theoretical properties hold: information network decomposes total correlation as a whole, and equivalence of the sum of in-coming and out-going information flows at any node in the network. These properties are stated in Theorem \ref{thm-TotalCorrelationDecompositionNVariables}, 
\ref{thm-InComingInformationFlows}, and \ref{thm-OutGoingInformationFlows}
%\ref{thm-LocalInformationFlow} 
whose proofs use Threorem \ref{thm-PartialExpansionInformationLattice}
as well as the three lemmas: identity of multivariate conditional mutual information (Lemma \ref{lem-IdentityMultivariateConditionalMutualInformation}), chain rule of multivariate conditional mutual information (Lemma 
\ref{lem-ChainRuleOfConditionalMutualInformation}), and decomposition of joint entropy (Lemma \ref{thm-JointEntropyDecomposition}).

%%%%%%%%%%%%%%%%%%%%%%%%%%%%%%%%%%%%%%%%%%%%%%%%%%%%%%%%%%%%%%%
%% TE Theorem 1: Conditional-TE-decomposition of total correlation
%%%%%%%%%%%%%%%%%%%%%%%%%%%%%%%%%%%%%%%%%%%%%%%%%%%%%%%%%%%%%%%
\begin{thm}[Decomposition of total correlation]
\label{thm-TotalCorrelationDecompositionNVariables}
In an $N$-variable system, total correlation 
consists of the sum of all the conditional transfer entropies and residual information.
%$N$-varaible total correlation (Equation \ref{eq-TotalCorrelationNVariables}) can be rewritten as follows.
\begin{eqnarray}
 C\left( \bar{X}_{ \mathcal{T} }^{\mathcal{N}} \right) 
=
\sum_{ t = 1}^{T}
\left[
R^{T}
+
\sum_{ i = 1 }^{N}
\left\{
R_{i}^{T}
+
\sum_{ j = 1 }^{ i - 1 }
\left( 
T^{T}_{ i \rightarrow j | \mathcal{ N } \setminus \{ i, j \} } + T^{T}_{ j \rightarrow i | \mathcal{ N } \setminus \{ i, j \} } 
+
R^{T}_{ i, j } 
\right)
\right\}
\right]
\label{eq-TotalCorrelationDecompositionNVariables}
\end{eqnarray}
where $
%\begin{equation}
 C( \bar{X}_{\mathcal{T}}^{\mathcal{N}} ) = \sum_{i=1}^{N}H( \bar{X}_{i}^{ \mathcal{T} }) - H( \bar{X}_{ \mathcal{T} }^{\mathcal{N}} )
%  \label{eq-TotalCorrelationNVariables}
%\end{equation}
$ is total correlation of the set of $N$ variables $\left\{ \bar{X}_{1}^{\mathcal{T}}, \bar{X}_{2}^{\mathcal{T}}, \hdots, \bar{X}_{N}^{\mathcal{T}} \right\}$.
\begin{proof}
 Each of the terms, transfer entropy and residual entropy, is rewritten in terms of the sum of multivariate conditional mutual information where
the spatio-temporal index of variables is of essential.
Thus, let us denote the sum of multivariate conditional mutual information satisties the inequality of indices as follows.
%\begin{equation}
%\sum_{ t_{s} } \tilde{I}( \bar{t}_{k}, t_{s} ) = 
% \sum_{i_{1}=1}^{t_{s_1}}
%\hdots \sum_{i_{|s|}=1}^{t_{ s_{|s|} }}
%\tilde{I}(t_{1}, \hdots t_{k}, i_{1}, \hdots, i_{ |s| }) 
%\end{equation}
%where $s \subset \{ s_{1}, s_{2}, \hdots, s_{ |s| }\}$ and $|s|$ is the size of the set $s$.
%where $\tilde{I}( t_{1}, t_{2}, \hdots, t_{N-1} ) = \tilde{I}( t_{1}, t_{2}, \hdots, t_{N-1}, \emptyset )$
%and
%$\tilde{I}( X_{1}^{ \bar{t}_{1} }, X_{2}^{ \bar{t}_{2} }, \hdots, X_{N}^{ \bar{t}_{N} } ) = \tilde{I}( t_{1}, t_{2}, \hdots, t_{N-1}, \emptyset )$
\begin{equation}
 \tilde{I}( X_{1}^{ \bar{t}_{1} }, X_{2}^{ \bar{t}_{2} }, \hdots, X_{N}^{ \bar{t}_{N} } ) 
= \sum_{s_{1} \subset \bar{t}_{1}}
\sum_{s_{2} \subset \bar{t}_{2}}
\hdots
\sum_{s_{N} \subset \bar{t}_{N}}
{I}( X_{1}^{ s_{1} }; X_{2}^{ s_{2} }; \hdots; X_{N}^{ s_{N} } | 
         X_{1}^{\bar{s}_{1}\setminus s_{1}}, X_{2}^{\bar{s}_{2}\setminus s_{2}}, \hdots, X_{N}^{\bar{s}_{N}\setminus s_{N}} ) 
\end{equation}
and
\begin{eqnarray}
 \tilde{I}\left( X_{1}^{ \bar{t}_{1} }, %X_{2}^{ \bar{t}_{2} }, 
 \hdots, X_{N}^{ \bar{t}_{N} } | 
\delta\left( {X}_{ \mathcal{N} }^{ \mathcal{T} } \right) \right) 
= \sum_{s_{1} \subset \bar{t}_{1}}
%\sum_{s_{2} \subset \bar{t}_{2}}
\hdots
\sum_{s_{N} \subset \bar{t}_{N}}
\delta\left( {X}_{ \mathcal{N} }^{ \mathcal{T} } \right)
{I}\left( X_{1}^{ s_{1} }; %X_{2}^{ s_{2} }; 
\hdots; X_{N}^{ s_{N} } | 
         X_{1}^{\bar{s}_{1}\setminus s_{1}}, X_{2}^{\bar{s}_{2}\setminus s_{2}}, \hdots, X_{N}^{\bar{s}_{N}\setminus s_{N}} \right) 
\end{eqnarray}
where $\delta( {X}_{ \mathcal{N} }^{ \mathcal{T} } )$ is the delta function which is 1 if the condition on the variables 
${X}_{ \mathcal{N} }^{ \mathcal{T} }$ is satisfied, zero otherwise.
Using the information lattice with the delta function, the transfer entropy is written as follows.
\begin{equation}
 T_{ j \rightarrow i | \mathcal{N} \setminus \{ i, j \}}^{T}
= \sum_{k=0}^{N-2} (-1)^{k+1} \sum_{ s \subset S( \mathcal{N}, k )}
 \tilde{I}( X_{i}^{\bar{t}_{i}}, X_{i}^{\bar{t}_{j}}, X_{ s_{1} }^{ \bar{t}_{s_1} }, \hdots, X_{ s_k }^{ \bar{t}_{s_k} } 
| t_{i} > t_{j} > \max( t_{ s_{1} }, t_{ s_{2} }, \hdots, t_{ s_{k} } ) ) 
\end{equation}
Likewise,
\begin{equation}
 R_{ i j | \mathcal{N} \setminus \{ i, j \}}^{T}
= \sum_{k=0}^{N-2} (-1)^{k+1} \sum_{ s \subset S( \mathcal{N}, k )}
 \tilde{I}( X_{i}^{\bar{t}_{i}}, X_{j}^{\bar{t}_{j}}, X_{ s_{1} }^{ \bar{t}_{s_1} }, \hdots, X_{ s_k }^{ \bar{t}_{s_k} } 
| t_{i} = t_{j} > \max( t_{ s_{1} }, t_{ s_{2} }, \hdots, t_{ s_{k} } ) ) 
\end{equation}
and
\begin{equation}
 R_{ i | \mathcal{N} \setminus \{ i \}}^{T}
= \sum_{k=0}^{N-1} (-1)^{k} \sum_{ s \subset S( \mathcal{N}, k )}
 \tilde{I}( X_{i}^{\bar{t}_{i}}, X_{ s_{1} }^{ \bar{t}_{s_1} }, \hdots, X_{ s_k }^{ \bar{t}_{s_k} } 
| t_{i} > \max( t_{ s_{1} }, \hdots, t_{ s_{k} } ) ) 
- \sum_{ j \subset \mathcal{N} \setminus i } T_{ j \rightarrow i | \mathcal{N} \setminus \{ i, j \}}
\label{eq-ResidualEntropySingleInformationLattice}
\end{equation}

\begin{equation}
 R^{T}
= \sum_{k=0}^{N-1} (-1)^{k} \sum_{ s \subset S( \mathcal{N}, k )}
 \tilde{I}( X_{ s_{1} }^{ \bar{t}_{s_1} }, \hdots, X_{ s_k }^{ \bar{t}_{s_k} } 
| t_{ s_{1} } = t_{ s_{2} } = \hdots = t_{ s_{k} } ) 
\end{equation}

Therefore, the right hand side of Equation \ref{eq-TotalCorrelationDecompositionNVariables} is as follows. 
\begin{eqnarray}
 \sum_{ k = 1}(-1)^{k+1} \sum_{s \subset S( \mathcal{N}, k ) } \tilde{I}( X_{ s }^{ \mathcal{T} } ) 
= C ( X_{ \mathcal{N} }^{ \mathcal{T} } ) 
\end{eqnarray}

\end{proof}
\end{thm}
%%%%%%%%%%%%%%%%%%%%%%%%%%%%%%%%%%%%%%%%%%%%%%%%%%%%%%%%%%%%%%%
%% TE Theorem 2: Local relationship of multivariate transfer entropy
%%%%%%%%%%%%%%%%%%%%%%%%%%%%%%%%%%%%%%%%%%%%%%%%%%%%%%%%%%%%%%%
%The second theorem states on two identities among free entropy, entropy rate, residual information, and conditional transfer entropy which guarantees equivalence between the sum of in-coming and out-going information flows at an arbitrary edge in the $N$-variable information network (Figure \ref{fig-LocalInformationFlow}).
\begin{thm}[In-coming information flow]
Entropy rate of variable $i$ consists of free entropy of variable $i$ and sum of in-coming conditional transfer entropy to variable $i$ as follows.
\label{thm-InComingInformationFlows}
\begin{eqnarray}
H_{i}^{T}
=
 F_{i}^{T} + \sum_{ j \subset \mathcal{N} \setminus i }
\left(
 T_{ j \rightarrow i | \mathcal{ N } \setminus \{ i, j \} }^{T} 
 + R_{ i | \mathcal{N} \setminus \{ i \}}^{T}
\right) 
\label{eq-InComingEntropyRate}
\end{eqnarray}

\begin{proof}
In terms of the information lattice, the difference of entropy rate from free entropy is rewritten as follows.
\begin{equation}
H_{i}^{T} - F_{i}^{T}
= \sum_{k=0}^{N-1} (-1)^{k+1} \sum_{ s \subset S( \mathcal{N}, k )}
 \tilde{I}( X_{i}^{\bar{t}_{i}}, X_{ s_{1} }^{ \bar{t}_{s_1} }, \hdots, X_{ s_k }^{ \bar{t}_{s_k} } 
| t_{i} > \max( t_{ s_{1} }, t_{ s_{2} }, \hdots, t_{ s_{k} } ) ) 
\end{equation}
Therefore, according to Equation \ref{eq-ResidualEntropySingleInformationLattice}, 
Equation \ref{eq-InComingEntropyRate} is proven.

\end{proof}
\end{thm}

%%%%%%%%%%%%%%%%%%%%%%%%%%%%%%%%%%%%%%%%%%%%%%%%%%%%%%%%%%%%%%%%%%%%%%%%%%%%%%%%%%%%%%%%%%%
%%%% Theorem 4
%%%%%%%%%%%%%%%%%%%%%%%%%%%%%%%%%%%%%%%%%%%%%%%%%%%%%%%%%%%%%%%%%%%%%%%%%%%%%%%%%%%%%%%%%%%
\begin{thm}[Out-going information flows]
\label{thm-OutGoingInformationFlows}
Entropy rate of variable $i$ can be locally decomposed with the sum of all the in-coming and out-going conditional transfer entropy and residual entropy.
\begin{equation}
H_{i}^{t} = H\left( X_{i}^{t} | X_{ i }^{\bar{t} \setminus t} \right)
\ge \sum_{ j \subset \mathcal{N} \setminus i } T_{ i \rightarrow j | \mathcal{N} \setminus i}
 \label{eq-OutGoingTransferEntropy}
\end{equation}

\begin{proof}
In terms of the information lattice, the difference of entropy rate from free entropy is rewritten as follows.
\begin{eqnarray}
H\left( X_{i}^{t} | X_{ \mathcal{N} }^{\bar{t} \setminus t}, X_{ \mathcal{N} \setminus i }^{ \mathcal{N} \setminus \bar{t} } \right)
=
\sum_{ j \subset \mathcal{N} \setminus i }
\sum_{k=0}^{N-2}(-1)^{ k }
\sum_{ s \subset S( \mathcal{N} \setminus \left\{ i, j \right\}, k )}
\tilde{I}( X_{i}^{\bar{t}_{i}}, X_{j}^{\bar{t}_{j}}, X_{ s_{1} }^{ \bar{t}_{s_1} }, \hdots, X_{ s_k }^{ \bar{t}_{s_k} } 
| t = t_{i}  ) ) 
\end{eqnarray}
where each of terms given a pair of indice $i$ and $j$ is non-negative.

On the other hand,
the sum of out-going information flows $\sum_{ j \subset \mathcal{N} \setminus i } T_{ i \rightarrow j | \mathcal{N} \setminus i}$ is 
\begin{eqnarray}
\sum_{ j \subset \mathcal{N} \setminus i } T_{ i \rightarrow j | \mathcal{N} \setminus i}
= \sum_{ j \subset \mathcal{N} \setminus i }
\sum_{k=0}^{N-2}(-1)^{ k }
\sum_{ s \subset S( \mathcal{N} \setminus \left\{ i, j \right\}, k )}
% \Big\{ 
\tilde{I}( X_{i}^{\bar{t}_{i}}, X_{j}^{\bar{t}_{j}}, X_{ s_{1} }^{ \bar{t}_{s_1} }, \hdots, X_{ s_k }^{ \bar{t}_{s_k} } 
| t_{j} > t = t_{i} > \max( t_{ s_{1} }, \hdots, t_{ s_{k} } ) ) 
\end{eqnarray}
Since the sum of out-going information flows is a subset of 
$H\left( X_{i}^{t} | X_{ \mathcal{N} }^{\bar{t} \setminus t}, X_{ \mathcal{N} \setminus i }^{ \mathcal{N} \setminus \bar{t} } \right)$, it gives the following inequality.
\begin{equation}
H\left( X_{i}^{t} | X_{ i }^{\bar{t} \setminus t} \right)
\ge
 H\left( X_{i}^{t} | X_{ \mathcal{N} }^{\bar{t} \setminus t}, X_{ \mathcal{N} \setminus i }^{ \mathcal{N} \setminus \bar{t} } \right)
\ge \sum_{ j \subset \mathcal{N} \setminus i } T_{ i \rightarrow j | \mathcal{N} \setminus i}
\end{equation}

\end{proof}
\end{thm}

\subsection{Visualization of the relationship between the information lattice and trivariate transfer entropy}
\begin{figure}[tb, clip]
   \begin{center}
    \includegraphics[width=1\linewidth, clip]{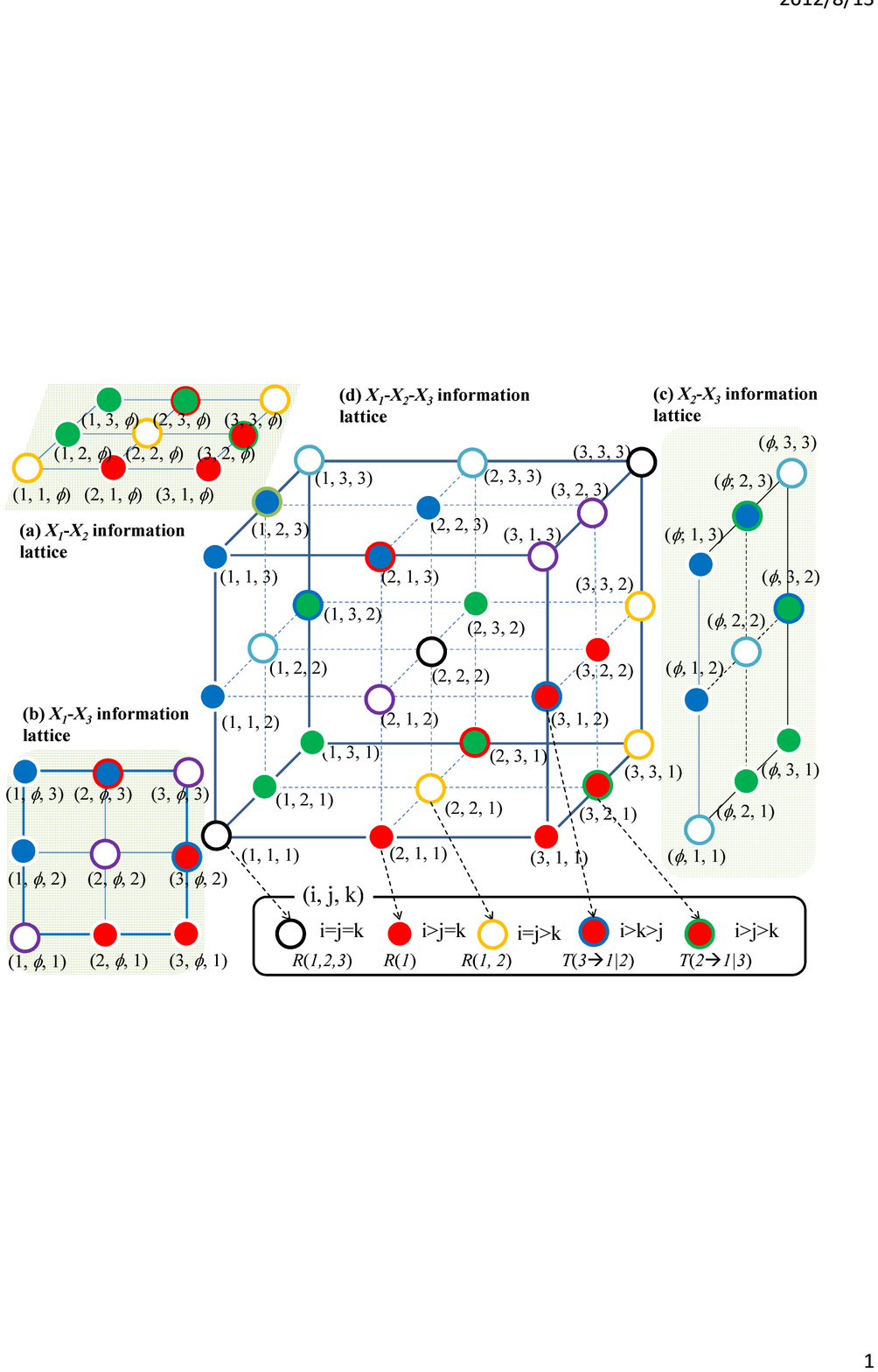}
    %\caption[]{Bidirectional information network between two variables $X$ and $Y$}  
    \caption{
The relationship between multivariate transfer entropy and 
information lattice for three groups of random variables 
$X_{1} = \{ X_{1}^{1}, X_{1}^{2}, X_{1}^{3}\}$, $X_{2} = \{ X_{2}^{1}, X_{2}^{2}, X_{2}^{3} \}$, and $X_{3} = \{ X_{3}^{1}, X_{3}^{2}, X_{3}^{3}\}$.
The coordinates of each node in the lattice indicates $(i, j, k)$ reflect the multivariate conditional 
mutual information $I(X_{1}^{i}, X_{2}^{j}, X_{3}^{k})$. In the special case, $(\emptyset, j, k)$, 
$(i, \emptyset, k)$, and $(i, j, \emptyset)$ respectively indicate 
$I(X_{2}^{j}, X_{3}^{k})$, $I(X_{1}^{i}, X_{3}^{k})$, $I(X_{1}^{i}, X_{2}^{j})$.
The red, green, and green nodes indicate the information flows to $X_{1}$, $X_{2}$, and $X_{3}$ respectively.
\label{fig-InformationLattice3D}}      
    \end{center}
  \end{figure}

As the minimum example, the information lattice of 
the trivariate system $ \{ X_{1}^{ \mathcal{T} }, X_{2}^{ \mathcal{T} }, X_{3}^{ \mathcal{T} }\}$ with $\mathcal{T} \subset \{ 1, 2, 3\}$ is depicted in Figure \ref{fig-InformationLattice3D}. The multivariate transfer entropy for the three variables is defined on the three bivariate information planes ((a) $X_{1}$-$X_{2}$, (b) $X_{1}$-$X_{3}$, (c) $X_{2}$-$X_{3}$) and the trivariate information cube ( (d) $X_{1}$-$X_{2}$-$X_{3}$ ). In Figure \ref{fig-InformationLattice3D}, each point on the information plane or cube is denoted by three dimensional coordinates 
indicating the time index of the three variables $(i, j, k) = \{ X_{1}^{i}, X_{2}^{j}, X_{3}^{k}\}$. In the information plane, $(i, j, \emptyset) = \{ X_{1}^{i}, X_{2}^{j}, \emptyset \}$ indicates the two dimensional coordinates.
The entropy rate of $i$ subtracted with its free entropy $(H_{1} - F_{1})$ is depicted with the circles
filled in red: $\{ - \tilde{I}( i, j, k | i > j \cap i > k ) + \tilde{I}( i, j, \emptyset | i > j ) + \tilde{I}( i, \emptyset, k | i > k ) \} $.
Likewise, $(H_{2} - F_{2})$ is depicted with the circles
filled in green: $\{ - \tilde{I}( i, j, k | j > i \cap j > k ) + \tilde{I}( i, j, \emptyset | j > i ) + \tilde{I}( \emptyset, j, k | j > k ) \} $, 
and $(H_{3} - F_{3})$ is also depicted with the circles
filled in green: $\{ - \tilde{I}( i, j, k | k > i \cap k > j ) + \tilde{I}( i, \emptyset, k | k > i ) + \tilde{I}( \emptyset, j, k | k > j ) \} $.
The transfer entropy is a subset of filled circles with colored outline, and 
its outline color of the circles indicate from which variable, in the same color code $\{ 1, 2, 3 \} = \{ R, G, B\}$, the information is transferred. 
Specifically, the coordinates $\{ \tilde{I}(3, \emptyset, 2 )- \tilde{I}(3, 1, 2) \}$ indicate
the transfer entropy from the variable 3 to variable 1 (the red-filled circle with blue outline), and 
the coordinates $\{ \tilde{I}(3, 2, \emptyset )- \tilde{I}(3, 2, 1) \}$ indicate
the transfer entropy from the variable 3 to variable 1 (the red-filled circle with green outline).
The sum of the transfer entropy from variable 2 and 3 to variable 1 is in-coming information flow to variable 1
(Theorem \ref{thm-InComingInformationFlows}),
and this is equal or less than $(H_{1} - F_{1})$ because the in-coming flow is a part of $(H_{1} - F_{1})$
and sum of information lattices given a pair of indices is non-negative (Corollary \ref{Cor-PairwiseInformationLattice})
and conditional mutual information is non-negative (Equation \ref{eq-ConditionalMutualInformation})).
The green and blue circles with red outline indicate out-going transfer entropy from variable 1 
($T_{1 \rightarrow 3} = \tilde{I}(2, \emptyset, 3 ) - \tilde{I}( 2, 1, 3 )$, $T_{1 \rightarrow 2} = \tilde{I}(2, 3, \emptyset)- \tilde{I}(2, 3, 1)$ ). The out-going transfer entropy is equal or less than
$\tilde{I}( 2, \emptyset, 1) - \tilde{I}( 2, 1, 1) \ge T_{1 \rightarrow 2} + T_{1 \rightarrow 3}$, thus,
it proves the trivariate special case of 
Theorem \ref{thm-OutGoingInformationFlows} by the inequality $H_{1} - F_{1} \ge T_{1 \rightarrow 2} + T_{1 \rightarrow 3}$.
The non-filled circles are called residual entropy, and each of these is classified by its index.
The sum of all the transfer entropies and residual entropies is equal to the total correlation of the three variables $C( X_{1}^{\mathcal{T}}, X_{2}^{\mathcal{T}}, X_{3}^{\mathcal{T}} ) = 
\sum_{i=1}^3\sum_{j=1}^{3}\tilde{I}( i, j, \emptyset )
+\sum_{i=1}^3\sum_{k=1}^{3}\tilde{I}( i, \emptyset, k )
+\sum_{j=1}^{3}\sum_{k=1}^{3}\tilde{I}( \emptyset, j, k )
-
\sum_{i=1}^3\sum_{j=1}^{3}\sum_{k=1}^{3}\tilde{I}( i, j, k )$, thus it visually proves Theorem \ref{thm-TotalCorrelationDecompositionNVariables} in the trivariate special case.

\bibliographystyle{pnas}
\bibliography{GTEReferences}